\documentclass[prd,showpacs,showkeys,floatfix,amsmath,amssymb,floatfix,english]{revtex4}

\usepackage{amsfonts}
\usepackage{subfig}
\usepackage{dcolumn}
\usepackage{natbib}
\usepackage{bm}
\usepackage{booktabs}
\usepackage{slashed}

\usepackage[T1]{fontenc}
\usepackage[utf8]{inputenc}
\usepackage{babel}
\usepackage{units}
\usepackage{amsmath}
\usepackage{esint}
\usepackage{multirow}
\usepackage{graphicx}
\usepackage[normalem]{ulem}
\usepackage[colorlinks, citecolor=blue,anchorcolor=red,menucolor=red, linkcolor=red,filecolor=red,runcolor=red,urlcolor=blue,frenchlinks=red]{hyperref}

\usepackage{ulem}
\usepackage{color}

\makeatletter
\newcommand{\figcaption}{\def\@captype{figure}\caption}
\newcommand{\tabcaption}{\def\@captype{table}\caption}

\newcommand{\Rmnum}[1]{\expandafter\@slowromancap\romannumeral #1@}

\newcommand{\tabincell}[2]{\begin{tabular}{@{}#1@{}}#2\end{tabular}}
\def\hlinewd#1{%
  \noalign{\ifnum0=`}\fi\hrule \@height #1 \futurelet
   \reserved@a\@xhline}
\makeatother

\begin{document}

\title{Fully open-flavor tetraquark states $bc\bar{q}\bar{s}$ and $sc\bar{q}\bar{b}$ with $J^{P}=0^{+},1^{+}$  }

\author{Qi-Nan Wang}
%\email{wangqn9@mail2.sysu.edu.cn}
\author{Wei Chen}
\email{chenwei29@mail.sysu.edu.cn}
\affiliation{School of Physics, Sun Yat-Sen University, Guangzhou 510275, China}

\begin{abstract}
We have studied the masses for fully open-flavor tetraquark states $bc\bar{q}\bar{s}$ and $sc\bar{q}\bar{b}$ with quantum numbers $J^{P}=0^{+},1^{+}$. We systematically construct all diquark-antiquark interpolating currents and calculate the two-point correlation functions and spectral densities in the framework of QCD sum rule method. Our calculations show that the masses are about $7.1-7.2$ GeV for the $bc\bar{q}\bar{s}$ tetraquark states and $7.0-7.1$ GeV for the $sc\bar{q}\bar{b}$ tetraquarks. The masses of $bc\bar{q}\bar{s}$ tetraquarks are below the thresholds of $\bar{B}_{s}D$ and $\bar{B}_{s}^{*}D$ final states for the scalar and axial-vector channels respectively. The $sc\bar{q}\bar{b}$ tetraquark states with $J^{P}=1^{+}$ lie below the $B_{c}^{+}K^{*}$ and $B_{s}^{*}D$ thresholds. Such low masses for these possible tetraquark states indicate that they can only decay via weak interaction and thus are very narrow and stable. 
\end{abstract}

%\date{\today}

\maketitle

\section{Introduction}
In the conventional quark model\cite{Jaffe:1976ig,GellMann:1964nj}, hadrons generally have two kinds of structures: a meson consisting of a quark and an antiquark, and a baryon consisting of three quarks. However, quantum chromodynamics (QCD) allows the existence of hadrons different from the above two structures, such as the tetraquarks, hadronic molecules, pentaquarks, hybrids and so on\cite{Chen:2016qju,2017-Lebed-p143-194,2018-Guo-p15004-15004,Liu:2019zoy}.

A compact tetraquak is composed of a diquark and an antidiquark, bounding by the color force among quarks and antiquarks. The light tetraquarks have been widely studied via different theoretical methods\cite{Chen:2007xr,Zhang:2006xp,Prelovsek:2008rf,Wallbott:2018lrl}. For the heavy quark sector, the hidden-charm/bottom $Q\bar{Q}q\bar{q}$ tetraquarks have been extensively investigated to interpret some observed XYZ states in various methods, such as the constituent quark models\cite{PhysRevD.98.094015,PhysRevD.94.014016,PhysRevD.94.074007}, meson exchange and scattering methods\cite{Liu:2017mrh,Ortega:2018cnm,Liu:2016kqx}, QCD sum rules\cite{2010-Nielsen-p41-83,2010-Chen-p105018-105018,2011-Chen-p34010-34010,2015-Chen-p54002-54002}, chromomagnetic interaction models\cite{Cui:2006mp,PhysRevD.79.077502}, etc. The doubly heavy tetraquark states $QQ\bar{q}\bar{q} $ have been studied to investigate the stability of tetraquarks\cite{Navarra:2007yw,2013-Du-p14003-14003}. In Ref. \cite{PhysRevD.89.054037,PhysRevD.99.014006,PhysRevD.99.054505}, the open-flavor heavy $bc\bar{q}\bar{q}$ tetraquark states have also been investigated, the results suggest that their masses may lie below the corresponding two-meson thresholds. 
In addition, such tetraquarks cannot decay via the annihilation channels and thus they will be very stable with narrow widths.

Comparing to the above several tetraquark configurations, the fully open-flavor tetraquarks $bc\bar{s}\bar{q}$ and $sc\bar{q}\bar{b}$ ($q=u, d$) are more exotic since they contain four valence quarks with totally different flavors. However, the studies of these tetraquarks have drawn much less interest to date. In Ref. \cite{Cui:2006mp}, the authors studied the masses of $qs\bar{c}\bar{b}$ and $qc\bar{s}\bar{b}$ tetraquarks by using the color-magnetic interaction with the flavor symmetry breaking corrections. Their results show that the masses of $qs\bar{c}\bar{b}$ and $qc\bar{s}\bar{b}$ tetraquark states are about 7.1 GeV and 7.2 GeV, which are lower than the corresponding two-meson S-wave thresholds. In the heavy quark symmetry, the mass of $bc\bar{q}\bar{s}$ tetraquark states with $J^P=1^+$ was also evaluated to be around 7445 MeV~\cite{Eichten:2017ffp}, which is about 163 MeV above the $D\bar B_s^\ast$ threshold and thus allows such decay channel via strong interaction. The above conflicting results from different phenomenological models are inspiring more theoretical studies for the existence of these fully open-flavor tetraquark states. In this paper, we shall study the mass spectra of the fully open-flavor $bc\bar{q}\bar{s}$ and $sc\bar{q}\bar{b}$ tetraquarks in the method of QCD sum rules \cite{Reinders:1984sr,Shifman:1978bx}.

This paper is organized as follows. In Sec.~\Rmnum{2}, we construct the interpolating tetraquark currents of the $bc\bar{q}\bar{s}$ and $sc\bar{q}\bar{b}$ systems with $J^{P}=0^{+},1^{+}$, respectively. In Sec.~\Rmnum{3}, we evaluate the correlation functions and spectral densities for these interpolating currents. The spectral densities will listed in the appendix because of their complicated form. We extract the masses for these tetraquarks by performing the QCD sum rule analyses in Sec.~\Rmnum{4}. The last section is a brief summary.

\section{Interpolating currents for the $bc\bar{q}\bar{s}$ and $sc\bar{q}\bar{b}$ tetraquark systems}
In this section, we construct the interpolating currents for $bc\bar{q}\bar{s}$ and $sc\bar{q}\bar{b}$ tetraquarks with $J^{P}=0^{+},1^{+}$. In general, there are five independent diquark fields,
$q_{a}^{T} C \gamma_{5} q_{b},~ q_{a}^{T} C q_{b},~ q_{a}^{T} C \gamma_{\mu} \gamma_{5} q_{b},~ q_{a}^{T} C \gamma_{\mu} q_{b},\text{and} ~q_{a}^{T} C \sigma_{\mu \nu} q_{b}$, where $q$ stands for quark field, $a,b$ are the color indices, $C$ denotes the charge conjugate operator, and $T$ represents the transpose of the quark fields.
The $q_{a}^{T} C \gamma_{5} q_{b}$ and $q_{a}^{T} C \gamma_{\mu} q_{b}$ are S-wave operators while $ q_{a}^{T} C q_{b}$ and $q_{a}^{T} C \gamma_{\mu} \gamma_{5} q_{b}$ are P-wave operators.
The $q_{a}^{T} C \sigma_{\mu \nu} q_{b}$ contains both S-wave and P-wave operators according to its different components.
To study the lowest lying $bc\bar{q}\bar{s}$ and $sc\bar{q}\bar{b}$ tetraquark states, we use only S-wave diquarks and corresponding antidiquark fields to construct the tetraquark interpolating currents with quantum numbers $J^{P}=0^{+},1^{+}$.
For the $bc\bar{q}\bar{s}$ system, the scalar currents with $J^{P}=0^{+}$ can be written as
\begin{equation}
\begin{aligned}
 J_{1} &=b_{a}^{T} C \gamma_{5} c_{b}\left(\bar{q}_{a} \gamma_{5} C \bar{s}_{b}^{T}+\bar{q}_{b} \gamma_{5} C \bar{s}_{a}^{T}\right)\, ,\\
  J_{2} &=b_{a}^{T} C \gamma_{5} c_{b}\left(\bar{q}_{a} \gamma_{5} C \bar{s}_{b}^{T}-\bar{q}_{b} \gamma_{5} C \bar{s}_{a}^{T}\right)\, , \\
  J_{3} &=b_{a}^{T} C \gamma_{\mu} c_{b}\left(\bar{q}_{a} \gamma^{\mu} C \bar{s}_{b}^{T}+\bar{q}_{b} \gamma^{\mu} C \bar{s}_{a}^{T}\right)\, , \\
   J_{4} &=b_{a}^{T} C \gamma_{\mu} c_{b}\left(\bar{q}_{a} \gamma^{\mu} C \bar{s}_{b}^{T}-\bar{q}_{b} \gamma^{\mu} C \bar{s}_{a}^{T}\right)\, , 
\label{scalarcurrents_bcqs}
 \end{aligned}
\end{equation}
in which $q$ is light quark field (up or down). The color structure for the currents $J_{1}$ and $J_{3}$ are symmetric 
$\left[\mathbf{6}_{\mathbf{c}}\right]_{b c} \otimes\left[\overline{\mathbf{6}}_{\mathbf{c}}\right]_{\bar{q} \bar{s}}$,
while for the $J_{2}$ and $J_{4}$ are antisymmetric 
$\left[\overline{\mathbf{3}}_{\mathbf{c}}\right]_{b c}\otimes\left[\mathbf{3}_{\mathbf{c}}\right]_{\bar{q} \bar{s}}$.
The axial-vector currents with $J^{P}=1^{+}$ can be written as
\begin{equation}
\begin{aligned}
 J_{1\mu} &=b_{a}^{T} C \gamma_{\mu} c_{b}\left(\bar{q}_{a} \gamma_{5} C \bar{s}_{b}^{T}+\bar{q}_{b} \gamma_{5} C \bar{s}_{a}^{T}\right)\, , \\
  J_{2\mu} &=b_{a}^{T} C \gamma_{\mu} c_{b}\left(\bar{q}_{a} \gamma_{5} C \bar{s}_{b}^{T}-\bar{q}_{b} \gamma_{5} C \bar{s}_{a}^{T}\right)\, , \\
  J_{3\mu} &=b_{a}^{T} C \gamma_{5} c_{b}\left(\bar{q}_{a} \gamma^{\mu} C \bar{s}_{b}^{T}+\bar{q}_{b} \gamma^{\mu} C \bar{s}_{a}^{T}\right)\, , \\
   J_{4\mu} &=b_{a}^{T} C \gamma_{5} c_{b}\left(\bar{q}_{a} \gamma^{\mu} C \bar{s}_{b}^{T}-\bar{q}_{b} \gamma^{\mu} C \bar{s}_{a}^{T}\right)\, ,
   \label{axialvectorcurrents_bcqs}
    \end{aligned}
\end{equation}
where the currents $J_{1\mu}$ and $J_{3\mu}$ are color symmetric 
while the $J_{2\mu}$ and $J_{4\mu}$ are color antisymmetric.

For the $sc\bar{q}\bar{b}$ system, the currents with $J^{P}=0^{+}$ are
\begin{equation}
\begin{aligned}
 \eta_{1} &=s_{a}^{T} C \gamma_{5} c_{b}\left(\bar{q}_{a} \gamma_{5} C \bar{b}_{b}^{T}+\bar{q}_{b} \gamma_{5} C \bar{b}_{a}^{T}\right)\, , \\
 \eta_{2} &=s_{a}^{T} C \gamma_{5} c_{b}\left(\bar{q}_{a} \gamma_{5} C \bar{b}_{b}^{T}-\bar{q}_{b} \gamma_{5} C \bar{b}_{a}^{T}\right)\, , \\
  \eta_{3} &=s_{a}^{T} C \gamma_{\mu} c_{b}\left(\bar{q}_{a} \gamma^{\mu} C \bar{b}_{b}^{T}+\bar{q}_{b} \gamma^{\mu} C \bar{b}_{a}^{T}\right)\, , \\
   \eta_{4} &=s_{a}^{T} C \gamma_{\mu} c_{b}\left(\bar{q}_{a} \gamma^{\mu} C \bar{b}_{b}^{T}-\bar{q}_{b} \gamma^{\mu} C \bar{b}_{a}^{T}\right)\, , \label{scalarcurrents_scqb}
    \end{aligned}
\end{equation}
where the currents $\eta_{1}$ and $\eta_{3}$ are color symmetric with 
$\left[\mathbf{6}_{\mathbf{c}}\right]_{s c} \otimes\left[\overline{\mathbf{6}}_{\mathbf{c}}\right]_{\bar{q} \bar{b}}$,
while the $\eta_{2}$ and $\eta_{4}$ are color antisymmetric with 
$\left[\overline{\mathbf{3}}_{\mathbf{c}}\right]_{s c}\otimes\left[\mathbf{3}_{\mathbf{c}}\right]_{\bar{q} \bar{b}}$. 
The currents with $J^{P}=1^{+}$ are
\begin{equation}
\begin{aligned}
 \eta_{1\mu} &=s_{a}^{T} C \gamma_{\mu} c_{b}\left(\bar{q}_{a} \gamma_{5} C \bar{b}_{b}^{T}+\bar{q}_{b} \gamma_{5} C \bar{b}_{a}^{T}\right)\, , \\
 \eta_{2\mu} &=s_{a}^{T} C \gamma_{\mu} c_{b}\left(\bar{q}_{a} \gamma_{5} C \bar{b}_{b}^{T}-\bar{q}_{b} \gamma_{5} C \bar{b}_{a}^{T}\right)\, , \\
  \eta_{3\mu} &=s_{a}^{T} C \gamma_{5} c_{b}\left(\bar{q}_{a} \gamma^{\mu} C \bar{b}_{b}^{T}+\bar{q}_{b} \gamma^{\mu} C \bar{b}_{a}^{T}\right)\, , \\
   \eta_{4\mu} &=s_{a}^{T} C \gamma_{5} c_{b}\left(\bar{q}_{a} \gamma^{\mu} C \bar{b}_{b}^{T}-\bar{q}_{b} \gamma^{\mu} C \bar{b}_{a}^{T}\right)\, , \label{axialvectorcurrents_scqb}
    \end{aligned}
\end{equation}
in which the currents $\eta_{1\mu}$ and $\eta_{3\mu}$ are color symmetric while 
the $\eta_{2\mu}$ and $\eta_{4\mu}$ are color antisymmetric.

\section{QCD sum rules}
In this section, we investigate the two-point correlation functions of the above scalar and axial-vector interpolating currents. For the scalar currents, the correlation function can be written as
\begin{equation}
\begin{aligned}
\Pi\left(p^{2}\right)&=i \int d^{4} x e^{i p \cdot x}\left\langle 0\left|T\left[J(x) J^{\dagger}(0)\right]\right| 0\right\rangle\, ,
\end{aligned}
\end{equation}
and for the axial-vector current
\begin{equation}
\begin{aligned}
 \Pi_{\mu \nu}\left(p^{2}\right) =i \int d^{4} x e^{i p \cdot x}\left\langle 0\left|T\left[J_{\mu}(x) J_{\nu}^{\dagger}(0)\right]\right| 0\right\rangle\, . 
 \label{CF_AV}
\end{aligned}
\end{equation}
The correlation function $\Pi_{\mu\nu} (p^{2})$ in Eq.~\eqref{CF_AV} can be expressed as 
\begin{equation}
\Pi_{\mu \nu}\left(p^{2}\right)=\left(\frac{p_{\mu} p_{\nu}}{p^{2}}-g_{\mu \nu}\right) \Pi_{1}\left(p^{2}\right)+\frac{p_{\mu} p_{\nu}}{p^{2}}\Pi_{0}\left(p^{2}\right)\, ,
\end{equation}
where $\Pi_{0}\left(p^{2}\right)$ and $\Pi_{1}\left(p^{2}\right)$ are the scalar and vector current polarization functions related to the spin-0 and spin-1 intermediate states, respectively.

 At the hadron level, the correlation function can be written through the dispersion relation
\begin{equation}
\Pi\left(p^{2}\right)=\frac{\left(p^{2}\right)^{N}}{\pi} \int_{(m_{b}+m_{c})^{2}}^{\infty} \frac{\operatorname{Im} \Pi(s)}{s^{N}\left(s-p^{2}-i \epsilon\right)} d s+\sum_{n=0}^{N-1} b_{n}\left(p^{2}\right)^{n}\, ,
\end{equation}
in which $b_n$ is the subtraction constant. In QCD sum rules, the imaginary part of the correlation function is defined as the spectral function 
\begin{equation}
\rho (s)=\frac{1}{\pi} \text{Im}\Pi(s)=f_{H}^{2}\delta(s-m_{H}^{2})+\text{QCD continuum and higher states}\, ,
\end{equation}
where the ``pole plus continuum parametrization" is adopted. The parameters $f_{H}$ and $m_{H}$ are the coupling constant and mass of the lowest-lying hadronic resonance $H$ respectively 
\begin{equation}
\begin{aligned}\langle 0|J| H\rangle &= f_{H}\, , \\
\left\langle 0\left|J_{\mu}\right| H\right\rangle &= f_{H} \epsilon_{\mu} \end{aligned}
\end{equation}
with the polarization vector $\epsilon_\mu$.

At the quark-gluon level, we can evaluate the correlation function $\Pi(p^{2})$ and spectral density $\rho(s)$ using the method of operator product expansion (OPE). To calculate the Wilson coefficients, we use the light quark propagator in coordinate space and heavy quark propagator in momentum space
\begin{equation}
\begin{aligned}
i S_{q}^{a b}(x)=& \frac{i \delta^{a b}}{2 \pi^{2} x^{4}} \hat{x}
+\frac{i}{32 \pi^{2}} \frac{\lambda_{a b}^{n}}{2} g_{s} G_{\mu \nu}^{n} \frac{1}{x^{2}}\left(\sigma^{\mu \nu} \hat{x}+\hat{x} \sigma^{\mu \nu}\right)
-\frac{\delta^{a b} x^{2}}{12}\left\langle\bar{q} g_{s} \sigma \cdot G q\right\rangle
-\frac{m_{q} \delta^{a b}}{4 \pi^{2} x^{2}} \\
&+\frac{i \delta^{a b} m_{q}(\bar{q} q)}{48} \hat{x}
-\frac{i m_{q}\left\langle\bar{q} g_{s} \sigma \cdot G q\right) \delta^{a b} x^{2} \hat{x}}{1152}\, , \\
 i S_{Q}^{a b}(p)=& \frac{i \delta^{a b}}{\hat{p}-m_{Q}}
 +\frac{i}{4} g_{s} \frac{\lambda_{a b}^{n}}{2} G_{\mu \nu}^{n} \frac{\sigma^{\mu \nu}\left(\hat{p}+m_{Q}\right)+\left(\hat{p}+m_{Q}\right) \sigma^{\mu \nu}}{12}
 +\frac{i \delta^{a b}}{12}\left\langle g_{s}^{2} G G\right\rangle m_{Q} \frac{p^{2}+m_{Q} \hat{p}}{(p^{2}-m_{Q}^{2})^{4}}\, , 
 \end{aligned}
\end{equation}
where $q$ represents $u$, $d$ or $s$ quark and $Q$ represents $c$ or $b$ quark. The superscripts $a, b$ are color indices and $\hat{x}=x^{\mu}\gamma_{\mu},\hat{p}=p^{\mu}\gamma_{\mu}$. In this work, we evaluate the Wilson coefficients up to dimension eight condensates at the leading order in $\alpha_s$. To improve the convergence of the OPE series and suppress the contributions from continuum and higher states region, one can perform the Borel transformation to the correlation function in both hadron and quark-gluon levels. The QCD sum rules are then established as
\begin{equation}
\mathcal{L}_{k}\left(s_{0}, M_{B}^{2}\right)=f_{H}^{2} m_{H}^{2 k} e^{-m_{H}^{2} / M_{B}^{2}}=\int_{(m_{b}+m_{c})^{2}}^{s_{0}} d s e^{-s / M_{B}^{2}} \rho(s) s^{k}\, ,
\label{Lk}
\end{equation}
where $M_B$ is the Borel mass introduced via the Borel transformation and $s_0$ the continuum threshold.
The lowest-lying hadron mass can be thus extracted via the following expression
\begin{equation}
\begin{aligned}
m_{H}\left(s_{0}, M_{B}^{2}\right)=&\sqrt{\frac{\mathcal{L}_{1}\left(s_{0}, M_{B}^{2}\right)}{\mathcal{L}_{0}\left(s_{0}, M_{B}^{2}\right)}}\, .
\end{aligned}
\end{equation}

\section{Numerical analysis}
In this section, we perform the QCD sum rule analyses for the $bc\bar{q}\bar{s}$ and $sc\bar{q}\bar{b}$ tetraquarks. We use the following values of quark masses and condensates\cite{Jamin:2001zr,Jamin:1998ra,Khodjamirian:2011ub,Tanabashi:2018oca,PhysRevD.99.054505}
\begin{equation}
\begin{array}{l}
{m_{u}(2 \mathrm{GeV})=(2.2_{-0.4}^{+0.5} ) \mathrm{MeV}}\ , \vspace{1ex}  \\
{m_{d}(2 \mathrm{GeV})=(4.7_{-0.3}^{+0.5}) \mathrm{MeV}}\, ,\vspace{1ex}  \\
{m_{q}(2 \mathrm{GeV})=(3.5_{-0.2}^{+0.5}) \mathrm{MeV}}\, ,\vspace{1ex}  \\
{m_{s}(2 \mathrm{GeV})=(95_{-3}^{+9}) \mathrm{MeV}}\, ,\vspace{1ex}  \\
{m_{c}\left(m_{c}\right)=(1.275 _{-0.035}^{+0.025}) \mathrm{GeV}}\, , \vspace{1ex} \\
{m_{b}\left(m_{b}\right)=(4.18 _{-0.03}^{+0.04}) \mathrm{GeV}}\, , \vspace{1ex} \\
\langle\bar{q} q\rangle=-(0.23 \pm 0.03)^{3} \mathrm{GeV}^{3}\, , \vspace{1ex} \\
{\left\langle\bar{q} g_{s} \sigma \cdot G q\right\rangle=- M_{0}^{2}\langle\bar{q} q\rangle}\, ,\vspace{1ex}  \\
{ M_{0}^{2}=(0.8 \pm 0.2) \mathrm{GeV}^{2}}\, , \vspace{1ex} \\
{\langle\bar{s} s\rangle /\langle\bar{q} q\rangle= 0.8 \pm 0.1}\, , \vspace{1ex} \\
{\left\langle g_{s}^{2} G G\right\rangle= (0.48\pm0.14) \mathrm{GeV}^{4}}\, ,
\end{array}
\end{equation}
in which the masses of $u,d,s$ are the current quark masses in the $\overline{MS}$ scheme at a scale $\mu = 2$ GeV. We consider the scale dependence of the charm and bottom quark masses at the leading order
\begin{equation}
\begin{aligned} m_{c}(\mu) &=\bar{m}_{c}\left(\frac{\alpha_{s}(\mu)}{\alpha_{s}\left(\bar{m}_{c}\right)}\right)^{12 / 25}\, ,
 \\ m_{b}(\mu) &=\bar{m}_{b}\left(\frac{\alpha_{s}(\mu)}{\alpha_{s}\left(\bar{m}_{b}\right)}\right)^{12 / 23}\, , 
 \end{aligned}
\end{equation}
where
\begin{equation}
\alpha_{s}(\mu)=\frac{\alpha_{s}\left(M_{\tau}\right)}{1+\frac{25 \alpha_{s}\left(M_{\tau}\right)}{12 \pi} \log \left(\frac{\mu^{2}}{M_{\tau}^{2}}\right)}, \quad \alpha_{s}\left(M_{\tau}\right)=0.33
\end{equation}
is determined by evolution from the $\tau$ mass using PDG values.

To obtain a stable sum rule, the working regions should also be determined, i.e, the continuum threshold $s_{0}$ and the Borel mass $M_{B}^{2}$. The threshold $s_{0}$ can be fixed by minimizing the variation of the hadronic masses $m_{H}$ with the Borel mass $M_{B}^{2}$. The Borel mass $M_{B}^{2}$ can be obtained by requiring the OPE convergence, which results in the lower bound of $M_{B}^{2}$, and a sufficient pole contribution, which results in the upper bound of $M_{B}^{2}$. Specific details of these procedures will be shown later. The pole contribution is defined as
\begin{equation}
\mathrm{PC}\left(s_{0}, M_{B}^{2}\right)=\frac{\mathcal{L}_{0}\left(s_{0}, M_{B}^{2}\right)}{\mathcal{L}_{0}\left(\infty, M_{B}^{2}\right)}\, ,
\end{equation}
in which $\mathcal{L}_{0}$ is defined in Eq.~(\ref{Lk}). 

\subsection{$bc\bar{q}\bar{s}$ systems}
\begin{figure}[h]
\centering
\includegraphics[width=10cm]{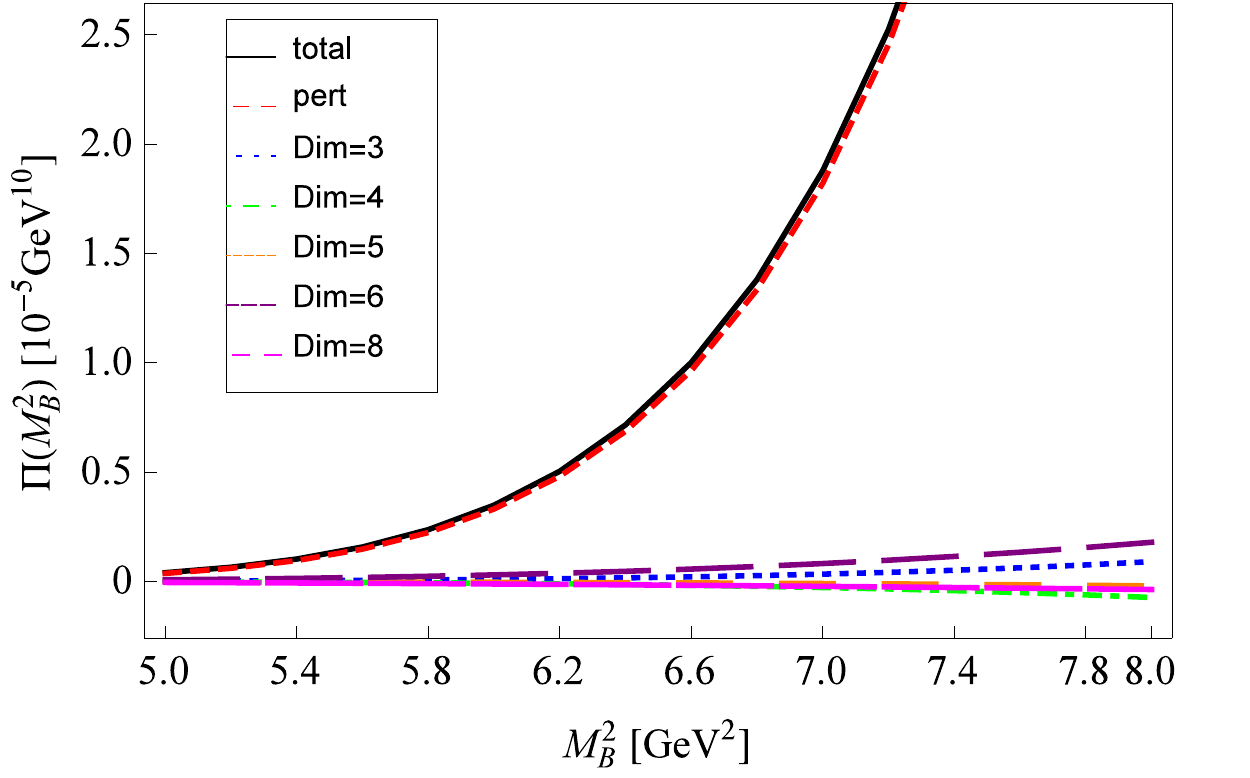}\\
\caption{OPE convergence for the current $J_{1}$ in the $J^{P}= 0^{+}$ $bc\bar{q}\bar{s}$ system}
\label{FigratioJ1}
\end{figure}

We firstly perform the QCD sum rule analyses for $bc\bar{q}\bar{s}$ tetraquarks. The spectral densities for the interpolating currents in Eqs.~\eqref{scalarcurrents_bcqs}--\eqref{axialvectorcurrents_bcqs} are calculated and listed in the appendix~\ref{spectral densities}.  For any interpolating current in the $bc\bar{q}\bar{s}$ system, contributions from the quark condensate $\langle\bar{q}q\rangle$ and quark-gluon mixed condensate $\langle\bar{q}g_{s}\sigma\cdot Gq\rangle$ are numerically small since they are proportional to the quark mass $m_{q}$ and $m_{s}$. The dominant nonperturbative contribution to the correlation function comes from the four-quark condensate $\langle\bar{q}q\rangle\langle\bar{s}s\rangle$. In Fig.\ref{FigratioJ1}, we take the scalar interpolating current $J_{1}$ as an example to present the contributions to correlation function from the perturbative and various condensate terms. To extract the output parameters, the Borel mass $M_{B}^{2}$ should be large enough to guarantee the convergence of OPE series. Here, we require that the four-quark condensate contribution be less than one-fifth of the
perturbative term. In Fig.\ref{FigratioJ1}, we can see that the convergence of OPE series can be ensured while $M_{B}^{2}\geq 5.4\text{GeV}^{2}$.
\begin{figure}[h]
\centering
  \includegraphics[width=8.5cm]{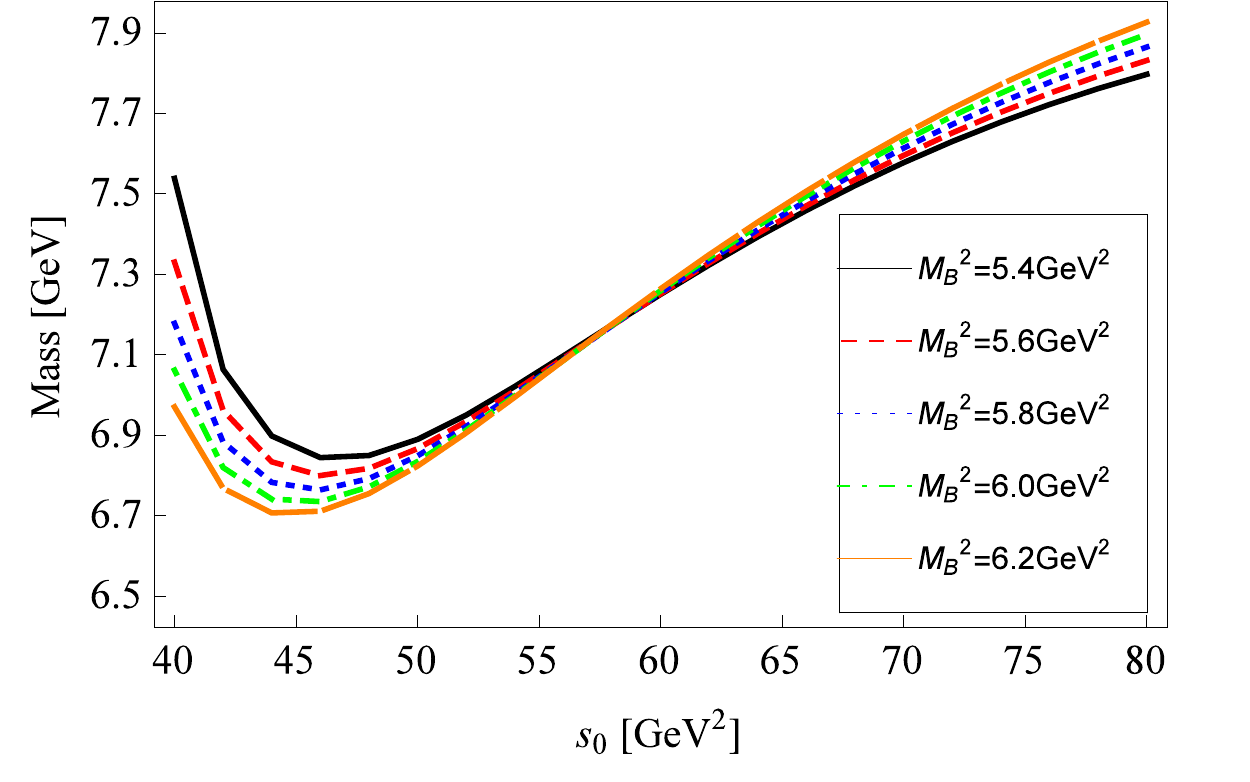}\quad
  \includegraphics[width=8.5cm]{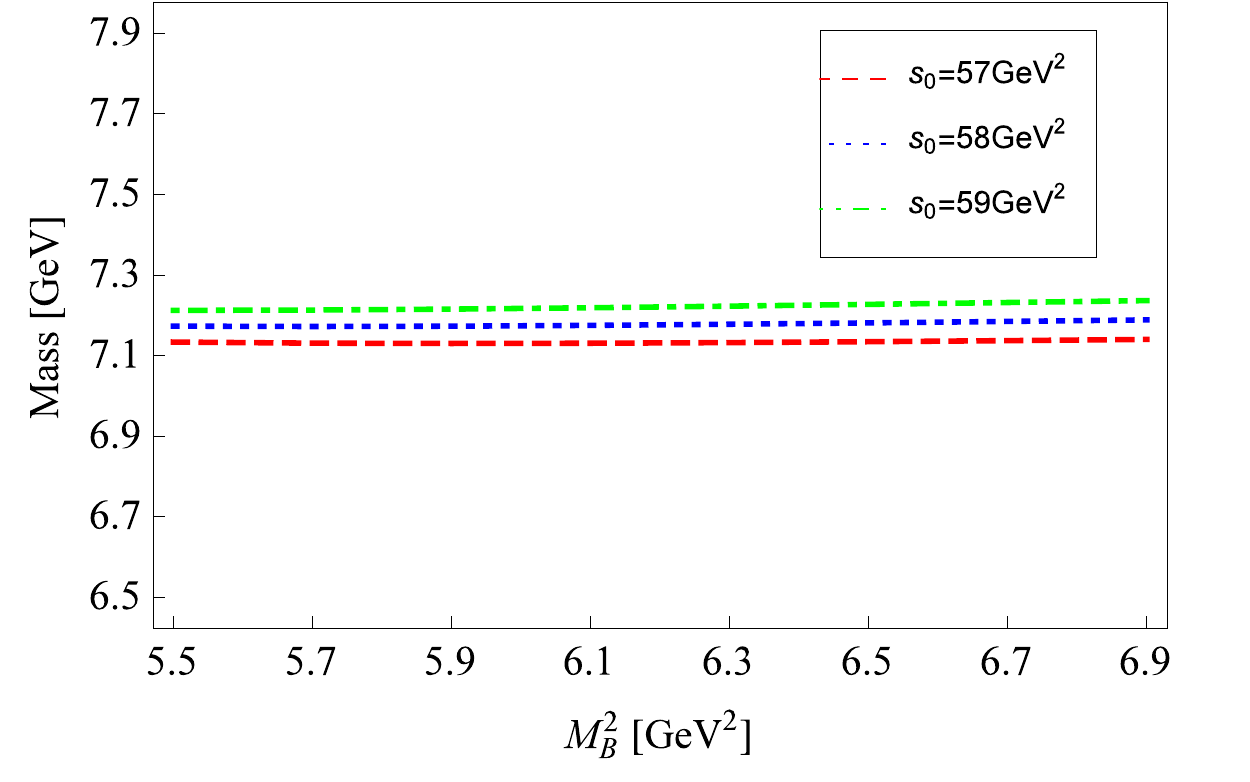}\\
\caption{Variation of $m_{H}$ with $s_{0}$ and $M_{B}^{2}$ corresponding to the current $J_{1}$ in the $J^{P}= 0^{+}$ $bc\bar{q}\bar{s}$ system}
  \label{FigMBandstaJ1}
\end{figure}

To get the upper bound of $M_{B}^{2}$, we need to fix the value of $s_{0}$ at first. As mentioned before, the output hadron mass $m_{H}$ should be irrelevant to $M_{B}^{2}$. In Fig.~\ref{FigMBandstaJ1}, we show the variations of $m_{H}$ with the threshold $s_{0}$ and Borel mass $M_{B}^{2}$. It is shown that the variation of $m_{H}$ with  $M_{B}^{2}$ minimizes around $s_{0}\sim 58~\text{GeV}^{2}$, which will result in the working region $56\leq s_0\leq 60~\text{GeV}^{2}$ for the scalar current $J_{1}$. Using this value of $s_{0}$, the upper bound of $M_{B}^{2}$ can be obtained by requiring the pole contribution be larger than 30\%. Finally, the working region of the Borel parameter for the scalar current $J_{1}$ can be determined to be $5.4\leq M_{B}^{2}\leq 5.8\text{GeV}^{2}$. We show the Borel curves in these regions in Fig.~\ref{FigMBandstaJ1} and extract the hadron mass to be $m_H=7.17\pm0.11$ GeV.  The errors come from the continuum threshold $s_{0}$, condensates $\langle\bar{q}q\rangle$ and  $\langle\bar{q}g_{s}\sigma\cdot Gq\rangle$, the heavy quark masses $m_{b}$ and $m_{c}$.  The errors from the Borel mass and the gluon condensate are small enough to be neglected.

For all other interpolating currents in Eqs.~\eqref{scalarcurrents_bcqs}--\eqref{axialvectorcurrents_bcqs}, 
we perform similar analyses and obtain the suitable working regions for the threshold $s_{0}$, Borel mass $M_{B}^{2}$, output hadron masses, pole contributions. We collect the numerical results in Table~\ref{resultone} for the scalar $bc\bar{q}\bar{s}$ tetraquarks and Table~\ref{resulttwo} for the axial-vector $bc\bar{q}\bar{s}$ tetraquarks. 
The last columns are the $S$-wave two-meson $\bar{B}_{s}D$ and $\bar{B}_{s}^\ast D$ thresholds for these possible tetraquark states. It is shown that both the scalar and axial-vector $bc\bar{q}\bar{s}$ tetraquarks lie below the corresponding two-meson thresholds, implying their stabilities against the strong interaction. 
\begin{table}[h]
\caption{The continuum threshold, Borel window, hadron mass and pole contribution for the $bc\bar{q}\bar{s}$ system with $J^{P} = 0^{+}$.}
\begin{center}
\label{resultone}
\begin{ruledtabular}
\begin{tabular}{cccccc }
  % after \\: \hline or \cline{col1-col2} \cline{col3-col4} ...
   Current&  $s_{0}$(\text{GeV$^{2}$}) & $M_{B}^{2}$(\text{GeV$^{2}$})  &$ m_{H}$(\text{GeV})  &PC(\%)  & \tabincell{c}{Two-meson threshold(GeV)}  \\ \hline 
  $J_{1}$ &        58$\pm$2            & 5.4$\sim$ 5.8                  & $7.17\pm0.11$        & 33.1   &             \vspace{1ex}  \\
  $J_{2}$ &        56$\pm$2            & 5.4$\sim$ 5.6                  & $7.04\pm0.13$        & 31.7   & 7.24     \vspace{1ex} \\
  $J_{3}$ &        57$\pm$2            & 5.4$\sim$ 5.6                  & $7.12\pm0.15$        & 32.7   &($\bar{B}_{s}D$)   \vspace{1ex} \\
  $J_{4}$ &        57$\pm$2            & 5.3$\sim$ 5.5                  & $7.11\pm0.20$        & 32.4   &\\
\end{tabular}
\end{ruledtabular}
\end{center}
\end{table}
\begin{table}[th]
\begin{center}
\caption{The continuum threshold, Borel window, hadron mass and pole contribution for the $bc\bar{q}\bar{s}$ system with $J^{P} = 1^{+}$.}
\label{resulttwo}
\begin{ruledtabular}
\begin{tabular}{cccccc}
  % after \\: \hline or \cline{col1-col2} \cline{col3-col4} ...
   Current&  $s_{0}$(\text{GeV$^{2}$}) & $M_{B}^{2}$(\text{GeV$^{2}$})    & $m_{H}$(\text{GeV})  & PC(\%)  & \tabincell{c}{Two-meson threshold(GeV)}  \\ \hline 
   $J_{1\mu}$ &        58$\pm$2            & 5.4$\sim$ 5.8                & $7.19\pm0.12$        & 33.1    &  \vspace{1ex} \\
    $J_{2\mu}$ &        57$\pm$2            & 5.3$\sim$ 5.7               & $7.10\pm0.11$        & 34.3    & 7.28   \vspace{1ex}\\
     $J_{3\mu}$ &        57$\pm$2            & 5.4$\sim$ 5.7              & $7.10\pm0.14$        & 33.9    &  ($\bar{B}_{s}^{*}D$)  \vspace{1ex}\\
     $J_{4\mu}$ &        58$\pm$2            & 5.4$\sim$ 5.9               & $7.16\pm0.13$       & 33.7    &    \\
\end{tabular}
\end{ruledtabular}
\end{center}
\end{table}

\subsection{$sc\bar{q}\bar{b}$ systems}
\begin{figure}[h]
\centering
\includegraphics[width=10cm]{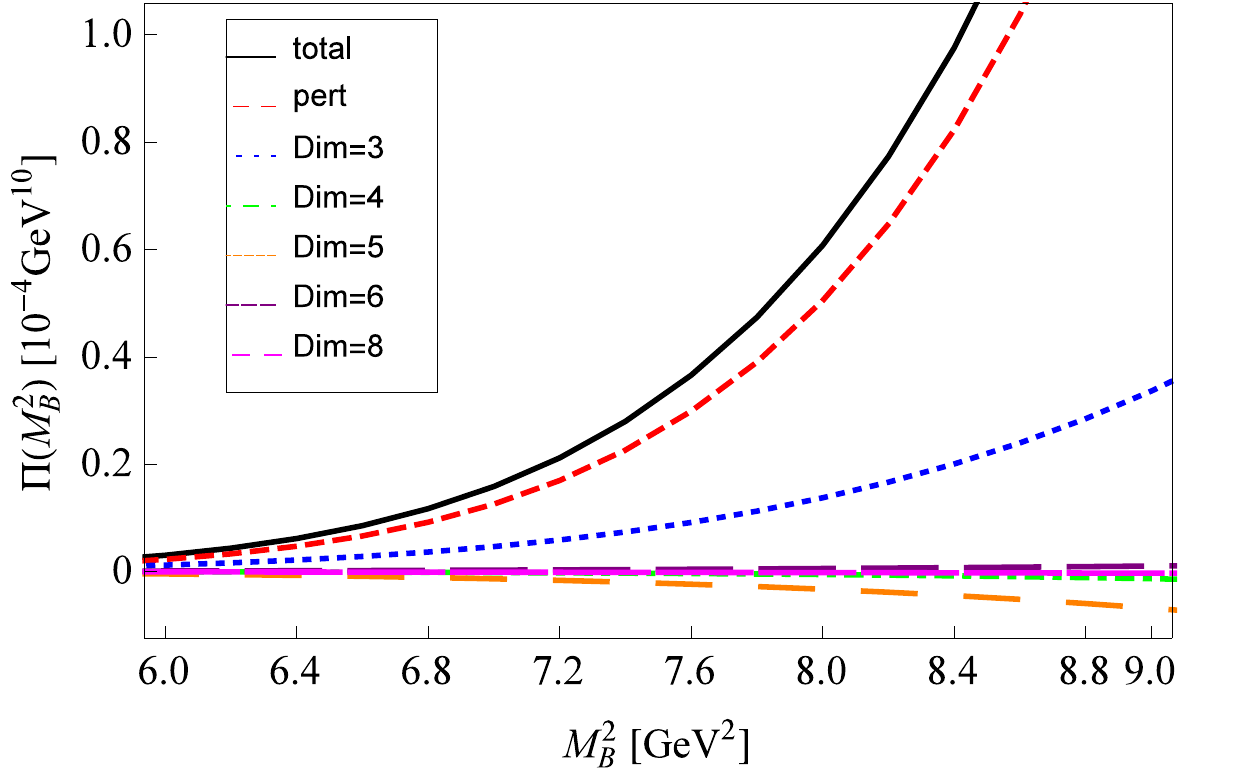}\\
\caption{The OPE behavior for the current $\eta_{1}$ in the $J^{P}= 0^{+}$ $sc\bar{q}\bar{b}$ system.}
\label{Figratioeta1}
\end{figure}

For the $sc\bar{q}\bar{b}$ systems, we calculate and list the correlation functions and spectral densities in the appendix~\ref{spectral densities} for all interpolating currents in Eqs.~\eqref{scalarcurrents_scqb}--\eqref{axialvectorcurrents_scqb}. Comparing to the $bc\bar{q}\bar{s}$ system, the OPE series behaviors are very different as shown in Fig.\ref{Figratioeta1} (for the scalar current $\eta_{1}$), where the contributions from the quark condensate $\langle\bar{q}q\rangle$ and quark-gluon mixed condensate $\langle\bar{q}g_{s}\sigma\cdot Gq\rangle$ are dominant while the contribution from the four-quark condensate $\langle\bar{q}q\rangle^{2}$ is relatively small. Such difference is due to the existence of $m_Q\langle\bar{q}q\rangle$ and $m_Q\langle\bar{q}g_{s}\sigma\cdot Gq\rangle$ in the $sc\bar{q}\bar{b}$ system. These terms are proportional to the heavy quark mass but absent in the OPE series of the $bc\bar{q}\bar{s}$ system. 
\begin{figure}[h]
\centering
  \includegraphics[width=8.5cm]{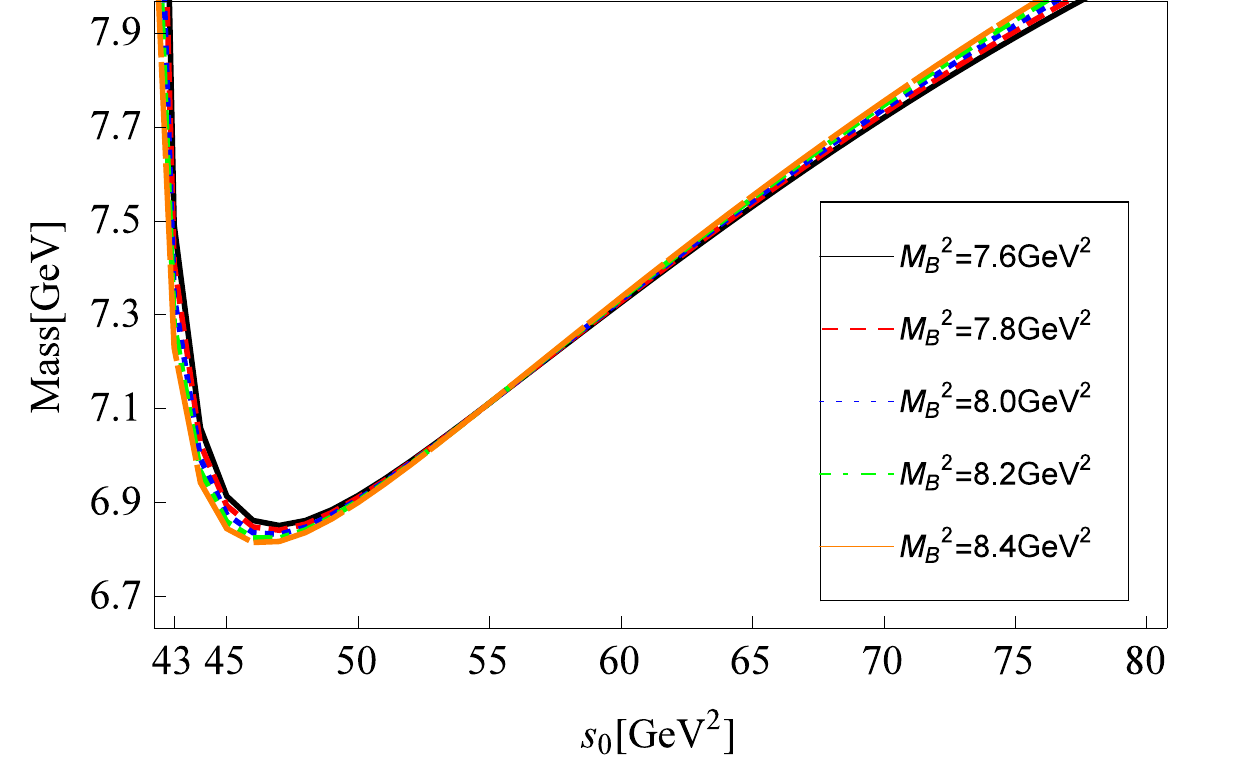}\quad
  \includegraphics[width=8.5cm]{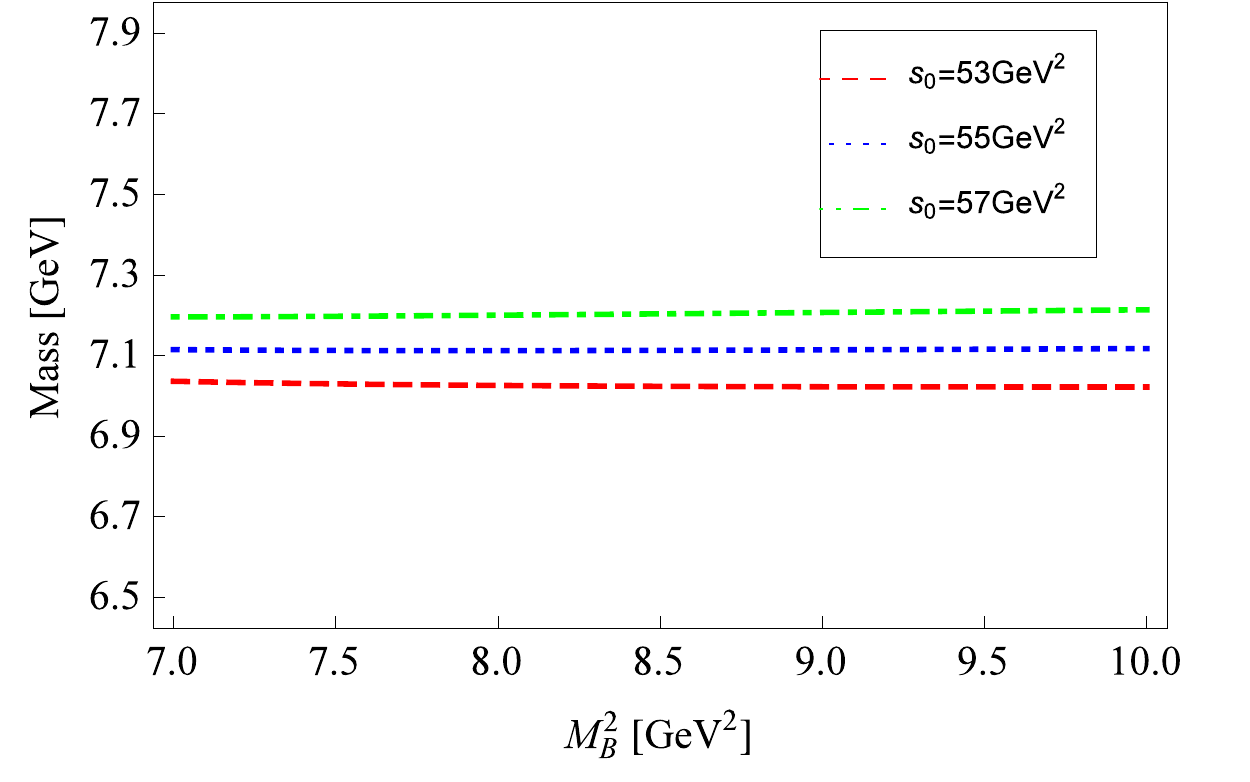}\\
\caption{Variation of $m_{H}$ with $s_{0}$ and $M_{B}^{2}$ corresponding to the current $\eta_{1}$ in the $J^{P}= 0^{+}$ $sc\bar{q}\bar{b}$ system}
  \label{FigMBandstaeta1}
\end{figure}

For the interpolating current $\eta_{1}$ with $J^P=0^+$, we perform the numerical analysis and find the suitable  working regions for the continuum threshold and Borel parameter are $55\leq s_0\leq 59~\text{GeV}^{2}$ and $7.6\leq M_B^2\leq 7.9~\text{GeV}^{2}$, respectively. We show the variations of $m_H$ with threshold $s_{0}$ and Borel mass $M_{B}^{2}$ in Fig.~\ref{FigMBandstaeta1}. Accordingly, the hadron mass can be extracted in these parameter regions. For all interpolating currents in the $sc\bar{q}\bar{b}$ systems, we can only establish reliable mass sum rules for  
$\eta_{1},\eta_{3},\eta_{4},\eta_{1\mu}$ and $\eta_{3\mu}$. We list the numerical results for the threshold $s_{0}$, Borel mass $M_{B}^{2}$, output masses, pole contributions  and the two-meson thresholds in Table~\ref{resultthree} for the scalar $sc\bar{q}\bar{b}$ system and Table~\ref{resultfour} for the axial-vector channel. In Table~\ref{resultthree}, the extracted masses for the scalar $sc\bar{q}\bar{b}$ tetraquarks are slightly below the $B_{s}D$ threshold while higher than the $B_{c}^{+}K$ threshold. However, the numerical results in Table~\ref{resultfour} show that the axial-vector $sc\bar{q}\bar{b}$ tetraquarks lie below both the $B_{s}^{*}D$ 
and $B_{c}^{+}K^{*}$ thresholds. 

\begin{table}[h!]
\begin{center}
\caption{The continuum threshold, Borel window, mass and pole contribution for the $sc\bar{q}\bar{b}$ system with $J^{P} = 0^{+}$.}
\label{resultthree}
\begin{ruledtabular}
\begin{tabular}{cccccc}
  % after \\: \hline or \cline{col1-col2} \cline{col3-col4} ...
   Current     &$s_{0}$(\text{GeV$^{2}$}) & $M_{B}^{2}$(\text{GeV$^{2}$})   & $m_{H}$(\text{GeV})  & PC(\%) & \tabincell{c}{Two-meson threshold(GeV)}  \\ \hline 
  $\eta_{1}$   & 55$\pm$2                 & 7.6$\sim$ 7.9                   & $7.11\pm0.11$          & 10.7   &    \vspace{1ex}\\
  $\eta_{2}$   & --                       & --                              & --                   & --     &6.77($B_{c}^{+}K$)   \vspace{1ex}\\
  $\eta_{3}$   & 54$\pm$2                 & 6.4$\sim$ 7.3                   & $7.02\pm0.12$          & 12.7   &7.24($B_{s}D$)  \vspace{1ex}\\
  $\eta_{4}$   & 56$\pm$2                 & 6.4$\sim$ 7.7                   & $7.13\pm0.12$        & 14.6   &\\
\end{tabular}
\end{ruledtabular}
\end{center}
\end{table}
\begin{table}[h!]
\caption{The continuum threshold, Borel window, mass and pole contribution for the $sc\bar{q}\bar{b}$ system with $J^{P} = 1^{+}$.}
\label{resultfour}
\begin{center}
\begin{ruledtabular}
\begin{tabular}{cccccc}
  % after \\: \hline or \cline{col1-col2} \cline{col3-col4} ...
   Current        & $s_{0}$(\text{GeV$^{2}$}) & $M_{B}^{2}$(\text{GeV$^{2}$})    & $m_{H}$(\text{GeV})   & PC(\%)   & \tabincell{c}{Two-meson threshold(GeV)}  \\ \hline 
 $\eta_{1\mu}$    & 55$\pm$2                  & 7.7$\sim$ 7.9                    & $7.10\pm0.12$         & 11.2    &  \vspace{1ex}\\
 $\eta_{2\mu}$    & --                        & --                               & --                    & --       &7.17($B_{c}^{+}K^{*}$)\vspace{1ex}\\
 $\eta_{3\mu}$    & 54$\pm$2                  & 7.6$\sim$ 7.8                    & $7.04\pm0.12$         & 10.4      &7.28($B_{s}^{*}D$) \vspace{1ex}\\
 $\eta_{4\mu}$    & --                        & --                               &--                     & --        &\\
\end{tabular}
\end{ruledtabular}
\end{center}
\end{table}

\section{Conclusion}
We have investigated the mass spectra for the fully open-flavored $bc\bar{q}\bar{s}$ and $sc\bar{q}\bar{b}$ tetraquark states in the framework of QCD sum rules. We construct the interpolating tetraquark currents with $J^{P}=0^{+}$ and $1^{+}$ and calculate their two-point correlation functions and spectral densities up to dimension eight condensates at the leading order of $\alpha_s$.

For the $bc\bar{q}\bar{s}$ tetraquark states, we find that the quark condensate $\langle\bar{q}q\rangle$ and quark-gluon mixed condensate $\langle\bar{q}g_{s}\sigma\cdot Gq\rangle$ are proportional to the light quark mass and thus numerically small. The dominant nonperturbative contribution to the correlation function comes from the four-quark condensate $\langle\bar{q}q\rangle\langle\bar{s}s\rangle$. The OPE series are very different for the $sc\bar{q}\bar{b}$ tetraquark systems, where the quark condensate and quark-gluon mixed condensate provide more important contribution than the four-quark condensate. Such difference leads to distinct behavior of the mass sum rules between the $bc\bar{q}\bar{s}$ and $sc\bar{q}\bar{b}$ tetraquark systems.

After the numerical analyses, we extract the masses around $7.1-7.2$ GeV for both the scalar and axial-vector $bc\bar{q}\bar{s}$ tetraquark states while $7.0-7.1$ GeV for the $sc\bar{q}\bar{b}$ tetraquarks. These results show that the masses of the $bc\bar{q}\bar{s}$ tetraquark states are below the $\bar{B}_{s}D$ and $\bar{B}_{s}^{*}D$ two-meson S-wave thresholds, which are consistent with the results from the color-magnetic interaction model~\cite{Cui:2006mp}. For the axial-vector $sc\bar{q}\bar{b}$ tetraquarks, their masses are also lower than the two-meson thresholds of $B_{c}^{+}K^{*}$ and $B_{s}^{*}D$ modes. Such results indicate that the two-meson strong decay modes are kinematically forbidden for these possible tetraquark states. They can only decay via the weak interaction if they do exist. For the scalar $sc\bar{q}\bar{b}$ tetraquark state, their decay to the $B_{s}D$ final states is also forbidden, but the $B_{c}^{+}K$ decay mode is allowed due to their slightly higher masses. However, such decay will be difficult since the low production rate of the $B_c^+$ meson. These stable teraquark states may be found in BelleII and LHCb in future.

\section*{ACKNOWLEDGMENTS}

This project is supported in part by the Chinese National Youth Thousand Talents Program.

\appendix
 \section{The spectral densities  \label{spectral densities}}
In this appendix, we list the spectral densities evaluated for the $bc\bar{q}\bar{s}$ and $sc\bar{q}\bar{b}$ tetraquark systems with $J^{P}=0^{+}$ and $1^{+}$. The spectral density includes the perturbative term, quark condensate, gluon condensate, quark-gluon mixed condensate, four-quark condensate  and dimension eight condensate  
\begin{equation}
\begin{aligned}
\rho(s)=\rho^{0}(s)+\rho^{3}(s)+\rho^{4}(s)+\rho^{5}(s)+\rho^{6}(s)+\rho^{8}(s)\, ,
\end{aligned}
\end{equation}
in which the superscripts stand for the dimension of various condensates.

\subsection{ The spectral densities for the $bc\bar{q}\bar{s}$ tetraquarks}
1.spectral densities for $J_{1}$:\\
\begin{equation}\nonumber
\rho_{J_{1}}^{0}(s)=
\int_{\alpha_{min}}^{\alpha_{max}} d \alpha \int_{\beta_{min}}^{\beta_{max}} d \beta
 \frac{(1-\alpha-\beta)^{2}(m_{b}^{2} \beta+m_{c}^{2} \alpha-\alpha \beta s)^{3}(m_{b}^{2} \beta+m_{c}^{2} \alpha-3 \alpha \beta s-2 m_{b} m_{c})} {256 \pi^{6} \alpha^{3} \beta^{3}}\, ,
\end{equation}
 \begin{equation}\nonumber
 \begin{aligned}
\rho_{J_{1}}^{3}(s)=&
\Big[m_{q}\langle\bar{q}q\rangle + m_{s}\langle\bar{s}s\rangle \Big]
\int_{\alpha_{min}}^{\alpha_{max}} d \alpha \int_{\beta_{min}}^{\beta_{max}} d \beta
 \frac{(m_{b}^{2} \beta+m_{c}^{2} \alpha-\alpha \beta s)(m_{b}^{2} \beta+m_{c}^{2} \alpha-2\alpha \beta s- m_{b} m_{c})} {16\pi^{4} \alpha \beta}\\
 &-\Big[m_{s}\langle\bar{q}q\rangle + m_{q}\langle\bar{s}s\rangle \Big ]
\int_{\alpha_{min}}^{\alpha_{max}} d \alpha \int_{\beta_{min}}^{\beta_{max}} d \beta
 \frac{(m_{b}^{2} \beta+m_{c}^{2} \alpha-\alpha \beta s)(m_{b}^{2} \beta+m_{c}^{2} \alpha-2\alpha \beta s- m_{b} m_{c})} {8 \pi^{4} \alpha \beta}\, ,
\end{aligned}
\end{equation}
\begin{equation}\nonumber
\begin{aligned}
\rho_{J_{1}}^{4a}(s)=
&\langle g_{s}^{2} G G\rangle \int_{\alpha_{min}}^{\alpha_{max}} d \alpha \int_{\beta_{min}}^{\beta_{max}} d \beta \frac{(1-\alpha-\beta)^{2}}{1536 \pi^{6}}
\Big[(2 m_{b}^{2} \beta+2 m_{c}^{2} \alpha-3 \alpha \beta s)\Big(\frac{m_{b}^{2}}{\alpha^{3}}+\frac{m_{c}^{2}}{\beta^{3}}\Big)\\
 &-\frac{m_{b} m_{c}}{\alpha \beta}\Big(\frac{4 m_{b}^{2} \beta+3 m_{c}^{2} \alpha-3 \alpha \beta s}{\alpha^{2}}+\frac{3 m_{b}^{2} \beta+4 m_{c}^{2} \alpha-3 \alpha \beta s}{\beta^{2}}\Big)\Big]\, ,
\end{aligned}
\end{equation}
\begin{equation}\nonumber
\begin{aligned}
\rho_{J_{1}}^{4b}(s)=
&-\langle g_{s}^{2} G G\rangle \int_{\alpha_{min}}^{\alpha_{max}} d \alpha \int_{\beta_{min}}^{\beta_{max}} d \beta \frac{( m_{b}^{2} \beta+ m_{c}^{2} \alpha- \alpha \beta s)}{1024 \pi^{6}}
\Big[\frac{( m_{b}^{2} \beta+ m_{c}^{2} \alpha-2 \alpha \beta s - m_{b}m_{c})}{\alpha \beta}\\
&+\frac{(1-\alpha-\beta)^{2}}{2\alpha^{2}\beta^{2}}( m_{b}^{2} \beta+ m_{c}^{2} \alpha-2 \alpha \beta s -2 m_{b}m_{c})\Big]\, ,
\end{aligned}
\end{equation}
\begin{equation}\nonumber
\begin{aligned}
\rho_{J_{1}}^{5}(s)=&
-\Big[m_{q}\langle\bar{s} g_{s}\sigma\cdot G s\rangle+m_{s}\langle\bar{q} g_{s}\sigma\cdot G q\rangle\Big]
\frac{s-(m_{b}-m_{c})^{2}}{64 \pi^{4}}
\sqrt{\left(1+\frac{m_{c}^{2}-m_{2}^{2}}{s}\right)^{2}-\frac{4 m_{b}^{2}}{s}}\\
&+\Big[m_{s}\langle\bar{s} g_{s}\sigma\cdot G s\rangle+m_{q}\langle\bar{q} g_{s}\sigma\cdot G q\rangle\Big]
\frac{s-(m_{b}-m_{c})^{2}}{192 \pi^{4}}
\sqrt{\left(1+\frac{m_{b}^{2}-m_{c}^{2}}{s}\right)^{2}-\frac{4 m_{b}^{2}}{s}}\, ,
\end{aligned}
\end{equation}
\begin{equation}\nonumber
\rho_{J_{1}}^{6}(s)=
\Big[\langle\bar{q} q\rangle \langle\bar{s} s\rangle\Big]
\frac{s-(m_{b}-m_{c})^{2}}{12 \pi^{2}}
\sqrt{\left(1+\frac{m_{b}^{2}-m_{c}^{2}}{s}\right)^{2}-\frac{4 m_{b}^{2}}{s}}\, ,
\end{equation}
\begin{equation}\nonumber
\begin{aligned}
\rho_{J_{1}}^{8}(s)=
&\Big[  \langle\bar{q} q\rangle \langle\bar{s}g_{s}\sigma\cdot G s\rangle+\langle\bar{s} s\rangle \langle\bar{q}g_{s}\sigma\cdot G q\rangle \Big]
 \int_{0}^{1} d \alpha \frac{1}{24 \pi^{2}}
 \Big\{\Big[\frac{m_{b} m_{c}^{3}+m_{c}^{4}\alpha+m_{b}^{2} m_{c}^{2}(1- \alpha)}{(1-\alpha)^{2}}\Big]\\
& \times\delta^{\prime}\Big[s-\frac{m_{b}^{2} \alpha+m_{c}^{2}(1-\alpha)}{\alpha(1-\alpha)}\Big]
+\frac{m_{b}^{2}(1- \alpha)+ 2m_{c}^{2}\alpha}{(1-\alpha)}
\delta\Big[s-\frac{m_{b}^{2}(1-\alpha)+m_{c}^{2} \alpha}{\alpha(1-\alpha)}\Big]\\
&-2\alpha H\Big[s-\frac{m_{b}^{2} (1-\alpha)+m_{c}^{2}\alpha}{\alpha(1-\alpha)}\Big]\Big\}\, ,
\end{aligned}
\end{equation}\\
where
\begin{equation}
\begin{aligned}
\alpha_{min }&=\frac{1}{2}\left[1+\frac{m_{b}^{2}-m_{c}^{2}}{s}-\sqrt{\left(1+\frac{m_{b}^{2}-m_{c}^{2}}{s}\right)^{2}-\frac{4 m_{b}^{2}}{s}}\right],\\
\alpha_{max }&=\frac{1}{2}\left[1+\frac{m_{b}^{2}-m_{c}^{2}}{s}+\sqrt{\left(1+\frac{m_{b}^{2}-m_{c}^{2}}{s}\right)^{2}-\frac{4 m_{b}^{2}}{s}}\right],\\
\beta_{min}&=\frac{\alpha m_{c}^{2}}{\alpha s-m_{b}^{2}},~~~~\beta_{max}=1-\alpha \, ,
 \end{aligned}
\end{equation}
 $\delta$ and $H$ denote the Dirac delta and Heavyside theta function, respectively.\\

2. Spectral densities for $J_{2}$:\\
$\rho_{J_{2}}^{0}(s)=\frac{1}{2}\rho_{J_{1}}^{0}(s)\, ,
~\rho_{J_{2}}^{3}(s)=\frac{1}{2}\rho_{J_{1}}^{3}(s)\, ,
~\rho_{J_{2}}^{4a}(s)=\frac{1}{2}\rho_{J_{1}}^{4a}(s)\, ,
~\rho_{J_{2}}^{4b}(s)=-\rho_{J_{1}}^{4b}(s)\, ,
~\rho_{J_{2}}^{5}(s)=\frac{1}{2}\rho_{J_{1}}^{5}(s)\, ,
~\rho_{J_{2}}^{6}(s)=\frac{1}{2}\rho_{J_{1}}^{6}(s)\, ,
~\rho_{J_{2}}^{8}(s)=\frac{1}{2}\rho_{J_{1}}^{8}(s)$\, .\\

3.Spectral densities for $J_{3}$:\\
 \begin{equation}\nonumber
\rho_{J_{3}}^{0}(s)=
\int_{\alpha_{min}}^{\alpha_{max}} d \alpha \int_{\beta_{min}}^{\beta_{max}} d \beta
 \frac{(1-\alpha-\beta)^{2}(m_{b}^{2} \beta+m_{c}^{2} \alpha-\alpha \beta s)^{3}(m_{b}^{2} \beta+m_{c}^{2} \alpha-3 \alpha \beta s- m_{b} m_{c})} {64 \pi^{6} \alpha^{3} \beta^{3}}\, ,
\end{equation}
\begin{equation}\nonumber
\begin{aligned}
\rho_{J_{3}}^{3}(s)=&
\Big[m_{q}\langle\bar{q}q\rangle + m_{s}\langle\bar{s}s\rangle \Big]
\int_{\alpha_{min}}^{\alpha_{max}} d \alpha \int_{\beta_{min}}^{\beta_{max}} d \beta
 \frac{(m_{b}^{2} \beta+m_{c}^{2} \alpha-\alpha \beta s)(2m_{b}^{2} \beta+ 2m_{c}^{2} \alpha-4\alpha \beta s- m_{b} m_{c})} {8\pi^{4} \alpha \beta}\\
 &-\Big[m_{s}\langle\bar{q}q\rangle + m_{q}\langle\bar{s}s\rangle \Big ]
\int_{\alpha_{min}}^{\alpha_{max}} d \alpha \int_{\beta_{min}}^{\beta_{max}} d \beta
 \frac{(m_{b}^{2} \beta+m_{c}^{2} \alpha-\alpha \beta s)(m_{b}^{2} \beta+m_{c}^{2} \alpha-2\alpha \beta s- 2m_{b} m_{c})} {4 \pi^{4} \alpha \beta}\, ,
\end{aligned}
\end{equation}
\begin{equation}\nonumber
\begin{aligned}
\rho_{J_{3}}^{4a}(s)=
&\langle g_{s}^{2} G G\rangle \int_{\alpha_{min}}^{\alpha_{max}} d \alpha \int_{\beta_{min}}^{\beta_{max}} d \beta \frac{(1-\alpha-\beta)^{2}}{384 \pi^{6}}
\Big[(2 m_{b}^{2} \beta+2 m_{c}^{2} \alpha-3 \alpha \beta s)\Big(\frac{m_{b}^{2}}{\alpha^{3}}+\frac{m_{c}^{2}}{\beta^{3}}\Big)\\
 &-\frac{m_{b} m_{c}}{2\alpha \beta}\Big(\frac{4 m_{b}^{2} \beta+3 m_{c}^{2} \alpha-3 \alpha \beta s}{\alpha^{2}}+\frac{3 m_{b}^{2} \beta+4 m_{c}^{2} \alpha-3 \alpha \beta s}{\beta^{2}}\Big)\Big]\, ,
\end{aligned}
\end{equation}
\begin{equation}\nonumber
\begin{aligned}
\rho_{J_{3}}^{4b}(s)=
&-\langle g_{s}^{2} G G\rangle \int_{\alpha_{min}}^{\alpha_{max}} d \alpha \int_{\beta_{min}}^{\beta_{max}} d \beta \frac{( m_{b}^{2} \beta+ m_{c}^{2} \alpha- \alpha \beta s)}{512 \pi^{6}}\frac{ m_{b}m_{c}}{\alpha \beta}\, ,
\end{aligned}
\end{equation}
\begin{equation}\nonumber
\begin{aligned}
\rho_{J_{3}}^{5}(s)=&
-\Big[m_{q}\langle\bar{s} g_{s}\sigma\cdot G s\rangle+m_{s}\langle\bar{q} g_{s}\sigma\cdot G q\rangle\Big]
\frac{s-(m_{b}-m_{c})^{2}+2m_{b}m_{c}}{32 \pi^{4}}
\sqrt{\left(1+\frac{m_{b}^{2}-m_{c}^{2}}{s}\right)^{2}-\frac{4 m_{b}^{2}}{s}}\\
&+\Big[m_{s}\langle\bar{s} g_{s}\sigma\cdot G s\rangle+m_{q}\langle\bar{q} g_{s}\sigma\cdot G q\rangle\Big]
\frac{s-(m_{b}-m_{c})^{2}-m_{b}m_{c}}{48 \pi^{4}}
\sqrt{\left(1+\frac{m_{b}^{2}-m_{c}^{2}}{s}\right)^{2}-\frac{4 m_{b}^{2}}{s}}\, ,
\end{aligned}
\end{equation}
\begin{equation}\nonumber
\rho_{J_{3}}^{6}(s)=
\Big[\langle\bar{q} q\rangle \langle\bar{s} s\rangle\Big]
\frac{s-(m_{b}-m_{c})^{2}+2m_{b}m_{c}}{6 \pi^{2}}
\sqrt{\left(1+\frac{m_{b}^{2}-m_{c}^{2}}{s}\right)^{2}-\frac{4 m_{b}^{2}}{s}}\, ,
\end{equation}
\begin{equation}\nonumber
\begin{aligned}
\rho_{J_{3}}^{8}(s)=
&\Big[  \langle\bar{q} q\rangle \langle\bar{s}g_{s}\sigma\cdot G s\rangle+\langle\bar{s} s\rangle \langle\bar{q}g_{s}\sigma\cdot G q\rangle \Big]
 \int_{0}^{1} d \alpha \frac{1}{12 \pi^{2}}
 \Big\{\Big[\frac{2 m_{b} m_{c}^{3}+m_{c}^{4}\alpha+m_{b}^{2} m_{c}^{2}(1- \alpha)}{(1-\alpha)^{2}}\Big]\\
& \times\delta^{\prime}\Big[s-\frac{m_{b}^{2} (1-\alpha)+m_{c}^{2}\alpha}{\alpha(1-\alpha)}\Big]
+\frac{m_{b}^{2}(1- \alpha)+ 2m_{c}^{2}\alpha}{(1-\alpha)}
\delta\Big[s-\frac{m_{b}^{2} (1-\alpha)+m_{c}^{2}\alpha}{\alpha(1-\alpha)}\Big]\\
&-2\alpha H\Big[s-\frac{m_{b}^{2} (1-\alpha)+m_{c}^{2}\alpha}{\alpha(1-\alpha)}\Big]\Big\}\, .
\end{aligned}
\end{equation}\\

4. Spectral densities for $J_{4}$:\\
$\rho_{J_{4}}^{0}(s)=\frac{1}{2}\rho_{J_{3}}^{0}(s)\, ,
~\rho_{J_{4}}^{3}(s)=\frac{1}{2}\rho_{J_{3}}^{3}(s)\, ,
~\rho_{J_{4}}^{4a}(s)=\frac{1}{2}\rho_{J_{3}}^{4a}(s)\, ,
~\rho_{J_{4}}^{4b}(s)=-\rho_{J_{3}}^{4b}(s)\, ,
~\rho_{J_{4}}^{5}(s)=\frac{1}{2}\rho_{J_{3}}^{5}(s)\, ,
~\rho_{J_{4}}^{6}(s)=\frac{1}{2}\rho_{J_{3}}^{6}(s)\, ,
~\rho_{J_{4}}^{8}(s)=\frac{1}{2}\rho_{J_{3}}^{8}(s)$\, .\\

5. Spectral densities for $J_{1\mu}$:\\
 \begin{equation}\nonumber
\rho_{J_{1\mu}}^{0}(s)=
\int_{\alpha_{min}}^{\alpha_{max}} d \alpha \int_{\beta_{min}}^{\beta_{max}} d \beta
 \frac{(1-\alpha-\beta)^{2}(m_{b}^{2} \beta+m_{c}^{2} \alpha-\alpha \beta s)^{3}(m_{b}^{2} \beta+m_{c}^{2} \alpha-5 \alpha \beta s- 4 m_{b} m_{c})} {512 \pi^{6} \alpha^{3} \beta^{3}}\, ,
\end{equation}
\begin{equation}\nonumber
 \begin{aligned}
\rho_{J_{1\mu}}^{3}(s)=&
\Big[m_{q}\langle\bar{q}q\rangle + m_{s}\langle\bar{s}s\rangle \Big]
\int_{\alpha_{min}}^{\alpha_{max}} d \alpha \int_{\beta_{min}}^{\beta_{max}} d \beta
 \frac{(m_{b}^{2} \beta+m_{c}^{2} \alpha-\alpha \beta s)(m_{b}^{2} \beta+ m_{c}^{2} \alpha-3\alpha \beta s- 2 m_{b} m_{c})} {32\pi^{4} \alpha \beta}\\
&-\Big[m_{s}\langle\bar{q}q\rangle + m_{q}\langle\bar{s}s\rangle \Big ]
\int_{\alpha_{min}}^{\alpha_{max}} d \alpha \int_{\beta_{min}}^{\beta_{max}} d \beta
 \frac{(m_{b}^{2} \beta+m_{c}^{2} \alpha-\alpha \beta s)(m_{b}^{2} \beta+m_{c}^{2} \alpha-3\alpha \beta s- 2 m_{b} m_{c})} {16 \pi^{4} \alpha \beta}\, ,
\end{aligned}
\end{equation}
\begin{equation}\nonumber
\begin{aligned}
\rho_{J_{1\mu}}^{4a}(s)=
&\langle g_{s}^{2} G G\rangle \int_{\alpha_{min}}^{\alpha_{max}} d \alpha \int_{\beta_{min}}^{\beta_{max}} d \beta \frac{(1-\alpha-\beta)^{2}}{1536 \pi^{6}}
\Big[( m_{b}^{2} \beta+ m_{c}^{2} \alpha-2 \alpha \beta s)\Big(\frac{m_{b}^{2}}{\alpha^{3}}+\frac{m_{c}^{2}}{\beta^{3}}\Big)\\
 &-\frac{m_{b} m_{c}}{\alpha \beta}\Big(\frac{4 m_{b}^{2} \beta+3 m_{c}^{2} \alpha-3 \alpha \beta s}{\alpha^{2}}+\frac{3 m_{b}^{2} \beta+4 m_{c}^{2} \alpha-3 \alpha \beta s}{\beta^{2}}\Big)\Big]\, ,
\end{aligned}
\end{equation}
\begin{equation}\nonumber
\begin{aligned}
\rho_{J_{1\mu}}^{4b}(s)=
&-\langle g_{s}^{2} G G\rangle \int_{\alpha_{min}}^{\alpha_{max}} d \alpha \int_{\beta_{min}}^{\beta_{max}} d \beta \frac{( m_{b}^{2} \beta+ m_{c}^{2} \alpha- \alpha \beta s)}{2048 \pi^{6}}
\Big[\frac{( m_{b}^{2} \beta+ m_{c}^{2} \alpha-3 \alpha \beta s -2 m_{b}m_{c})}{\alpha \beta}\\
&-\frac{(1-\alpha-\beta)^{2}}{6\alpha^{2}\beta^{2}}( 3m_{b}^{2} \beta+ 3 m_{c}^{2} \alpha-5 \alpha \beta s )\Big]\, ,
\end{aligned}
\end{equation}
\begin{equation}\nonumber
\begin{aligned}
\rho_{J_{1\mu}}^{5a}(s)=&
-\Big[m_{q}\langle\bar{s} g_{s}\sigma\cdot G s\rangle+m_{s}\langle\bar{q} g_{s}\sigma\cdot G q\rangle\Big]
\frac{[(m_{b}+m_{c})^{2}+2s][s-(m_{b}-m_{c})^{2}]}{192 \pi^{4}s}
\sqrt{\left(1+\frac{m_{b}^{2}-m_{c}^{2}}{s}\right)^{2}-\frac{4 m_{b}^{2}}{s}}\\
&+\Big[m_{s}\langle\bar{s} g_{s}\sigma\cdot G s\rangle+m_{q}\langle\bar{q} g_{s}\sigma\cdot G q\rangle\Big]
\frac{[(m_{b}+m_{c})^{2}+2s][s-(m_{b}-m_{c})^{2}]}{576 \pi^{4}s}
\sqrt{\left(1+\frac{m_{b}^{2}-m_{c}^{2}}{s}\right)^{2}-\frac{4 m_{b}^{2}}{s}}\, ,
\end{aligned}
\end{equation}
\begin{equation}\nonumber
\rho_{J_{1\mu}}^{6}(s)=
\Big[\langle\bar{q} q\rangle \langle\bar{s} s\rangle\Big]
\frac{[(m_{b}+m_{c})^{2}+2s][s-(m_{b}-m_{c})^{2}]}{36 \pi^{2}s}
\sqrt{\left(1+\frac{m_{b}^{2}-m_{c}^{2}}{s}\right)^{2}-\frac{4 m_{b}^{2}}{s}}\, ,
\end{equation}
\begin{equation}\nonumber
\begin{aligned}
\rho_{J_{1\mu}}^{8}(s)=
&\Big[  \langle\bar{q} q\rangle \langle\bar{s}g_{s}\sigma\cdot G s\rangle+\langle\bar{s} s\rangle \langle\bar{q}g_{s}\sigma\cdot G q\rangle \Big]
 \int_{0}^{1} d \alpha \frac{1}{24 \pi^{2}}
 \Big\{\Big[\frac{ m_{b} m_{c}^{3}+m_{c}^{4}\alpha+m_{b}^{2} m_{c}^{2}(1- \alpha)}{(1-\alpha)^{2}}\Big]\\
& \times\delta^{\prime}\Big[s-\frac{m_{b}^{2} (1-\alpha)+m_{c}^{2}\alpha}{\alpha(1-\alpha)}\Big]
+\frac{m_{b}^{2}(1- \alpha)+ m_{c}^{2}\alpha}{(1-\alpha)}
\delta\Big[s-\frac{m_{b}^{2} (1-\alpha)+m_{c}^{2}\alpha}{\alpha(1-\alpha)}\Big]\Big\}\, .
\end{aligned}
\end{equation}\\

6. Spectral densities for $J_{2\mu}$:\\
$\rho_{J_{2\mu}}^{0}(s)=\frac{1}{2}\rho_{J_{1\mu}}^{0}(s)\, ,
~\rho_{J_{2\mu}}^{3}(s)=\frac{1}{2}\rho_{J_{1\mu}}^{3}(s)\, ,
~\rho_{J_{2\mu}}^{4a}(s)=\frac{1}{2}\rho_{J_{1\mu}}^{4a}(s)\, ,
~\rho_{J_{2\mu}}^{4b}(s)=-\rho_{J_{1\mu}}^{4b}(s)\, ,
~\rho_{J_{2\mu}}^{5}(s)=\frac{1}{2}\rho_{J_{1\mu}}^{5}(s)\, ,
~\rho_{J_{2\mu}}^{6}(s)=\frac{1}{2}\rho_{J_{1\mu}}^{6}(s)\, ,
~\rho_{J_{2\mu}}^{8}(s)=\frac{1}{2}\rho_{J_{1\mu}}^{8}(s)$\, .\\

7.Spectral densities for $J_{3\mu}$:\\
\begin{equation}\nonumber
\begin{aligned}
\rho_{J_{3\mu}}^{0}(s)=&
\int_{\alpha_{min}}^{\alpha_{max}} d \alpha \int_{\beta_{min}}^{\beta_{max}} d \beta
 \frac{(1-\alpha-\beta)^{2}(m_{b}^{2} \beta+m_{c}^{2} \alpha-\alpha \beta s)^{3}} {1536 \pi^{6} \alpha^{3} \beta^{3}}
 \Big[6(m_{b}^{2} \beta+m_{c}^{2} \alpha-3 \alpha \beta s- 2 m_{b} m_{c})\\
 &-(1-\alpha-\beta)(3m_{b}^{2} \beta+3m_{c}^{2} \alpha-7 \alpha \beta s- 4 m_{b} m_{c})\Big]\, ,
 \end{aligned}
\end{equation}
\begin{equation}\nonumber
 \begin{aligned}
\rho_{J_{3\mu}}^{3}(s)=&
\Big[m_{q}\langle\bar{q}q\rangle + m_{s}\langle\bar{s}s\rangle \Big]
\int_{\alpha_{min}}^{\alpha_{max}} d \alpha \int_{\beta_{min}}^{\beta_{max}} d \beta
 \frac{(m_{b}^{2} \beta+m_{c}^{2} \alpha-\alpha \beta s)} {16 \pi^{4} \alpha \beta}
 \Big[(m_{b}^{2} \beta+m_{c}^{2} \alpha-2 \alpha \beta s- m_{b} m_{c})\\
 &-\frac{1}{2}(1-\alpha-\beta)(3m_{b}^{2} \beta+3m_{c}^{2} \alpha-5 \alpha \beta s- 2 m_{b} m_{c})\Big]\\
&-\Big[m_{s}\langle\bar{q}q\rangle + m_{q}\langle\bar{s}s\rangle \Big ]
\int_{\alpha_{min}}^{\alpha_{max}} d \alpha \int_{\beta_{min}}^{\beta_{max}} d \beta
 \frac{(m_{b}^{2} \beta+m_{c}^{2} \alpha-\alpha \beta s)} {8 \pi^{4} \alpha \beta}
 \Big[(m_{b}^{2} \beta+m_{c}^{2} \alpha-2 \alpha \beta s- m_{b} m_{c})\Big]\, ,
 \end{aligned}
\end{equation}
\begin{equation}\nonumber
\begin{aligned}
\rho_{J_{3\mu}}^{4a}(s)=
&\langle g_{s}^{2} G G\rangle \int_{\alpha_{min}}^{\alpha_{max}} d \alpha \int_{\beta_{min}}^{\beta_{max}} d \beta
\frac{(1-\alpha -\beta)^2 }{4608 \pi ^6}
\Big\{ \big[3 (2   m_{b}^{2} \beta+2 m_{c}^{2}  \alpha -3 \alpha  \beta  s)\\
&-(1-\alpha -\beta ) (3  m_{b}^{2} \beta +3  m_{c}^{2} \alpha-4 \alpha  \beta  s)\big]\Big(\frac{m_{b}^{2}}{\alpha ^3}+\frac{m_{c}^{2}}{\beta ^3}\Big) \\
&-\frac{ m_{b}  m_{c} (\alpha +\beta +2)}{\alpha  \beta } \Big(\frac{4 m_{b}^{2} \beta +3  m_{c}^{2} \alpha-3 \alpha  \beta  s}{\alpha ^2}+\frac{3 m_{b}^{2} \beta +4  m_{c}^{2} \alpha-3 \alpha  \beta  s}{\beta ^2}\Big)\Big\}\, ,
\end{aligned}
\end{equation}
\begin{equation}\nonumber
\begin{aligned}
\rho_{J_{3\mu}}^{4b}(s)=
&\langle g_{s}^{2} G G\rangle \int_{\alpha_{min}}^{\alpha_{max}} d \alpha \int_{\beta_{min}}^{\beta_{max}} d \beta \frac{( m_{b}^{2} \beta+ m_{c}^{2} \alpha- \alpha \beta s)}{12288 \pi^{6}}
\Big\{\frac{2}{\alpha \beta}\big[2 (m_{b}^{2} \beta+ m_{c}^{2}  \alpha -2 \alpha  \beta  s - m_{b}m_{c})\\
&+(1-\alpha -\beta ) (3  m_{b}^{2} \beta +3  m_{c}^{2} \alpha-5 \alpha  \beta  s  -2 m_{b}m_{c})\big]
-\frac{(1-\alpha-\beta)^{2}}{\alpha^{2}\beta^{2}}\big[6(m_{b}^{2} \beta+ m_{c}^{2}  \alpha -2 \alpha  \beta  s- 2m_{b}m_{c})\\
&-(1-\alpha -\beta ) (3  m_{b}^{2} \beta +3  m_{c}^{2} \alpha-5 \alpha  \beta  s- 4m_{b}m_{c})\big]\Big\}\, ,
\end{aligned}
\end{equation}
\begin{equation}\nonumber
\begin{aligned}
\rho_{J_{3\mu}}^{5a}(s)=&
-\Big[m_{q}\langle\bar{s} g_{s}\sigma\cdot G s\rangle+m_{s}\langle\bar{q} g_{s}\sigma\cdot G q\rangle\Big]
\frac{s-(m_{b}-m_{c})^{2}}{64 \pi^{4}}
\sqrt{\left(1+\frac{m_{b}^{2}-m_{c}^{2}}{s}\right)^{2}-\frac{4 m_{b}^{2}}{s}}\\
&+\Big[m_{s}\langle\bar{s} g_{s}\sigma\cdot G s\rangle+m_{q}\langle\bar{q} g_{s}\sigma\cdot G q\rangle\Big]
\frac{s-(m_{b}-m_{c})^{2}}{192 \pi^{4}}
\sqrt{\left(1+\frac{m_{b}^{2}-m_{c}^{2}}{s}\right)^{2}-\frac{4 m_{b}^{2}}{s}}\, ,
\end{aligned}
\end{equation}
\begin{equation}\nonumber
\rho_{J_{3\mu}}^{5b}(s)=
\Big[m_{s}\langle\bar{s} g_{s}\sigma\cdot G s\rangle+m_{q}\langle\bar{q} g_{s}\sigma\cdot G q\rangle\Big]\int_{\alpha_{min}}^{\alpha_{max}} d \alpha \int_{\beta_{min}}^{\beta_{max}} d \beta
 \frac{(3m_{b}^{2} \beta+3m_{c}^{2} \alpha-4\alpha \beta s-m_{b}m_{c})} {96 \pi^{4}}\, ,
\end{equation}
\begin{equation}\nonumber
\rho_{J_{3\mu}}^{6}(s)=
\Big[\langle\bar{q} q\rangle \langle\bar{s} s\rangle\Big]
\frac{s-(m_{b}-m_{c})^{2}}{12 \pi^{2}}
\sqrt{\left(1+\frac{m_{b}^{2}-m_{c}^{2}}{s}\right)^{2}-\frac{4 m_{b}^{2}}{s}}\, ,
\end{equation}
\begin{equation}\nonumber
\begin{aligned}
\rho_{J_{3\mu}}^{8}(s)=
&\Big[  \langle\bar{q} q\rangle \langle\bar{s}g_{s}\sigma\cdot G s\rangle+\langle\bar{s} s\rangle \langle\bar{q}g_{s}\sigma\cdot G q\rangle \Big]
 \int_{0}^{1} d \alpha \frac{1}{24 \pi^{2}}
 \Big\{\Big[\frac{ m_{b} m_{c}^{3}+m_{c}^{4}\alpha+m_{b}^{2} m_{c}^{2}(1- \alpha)}{(1-\alpha)^{2}}\Big]\\
& \times\delta^{\prime}\Big[s-\frac{m_{b}^{2} (1-\alpha)+m_{c}^{2}\alpha}{\alpha(1-\alpha)}\Big]
+\frac{m_{b}^{2}(1- \alpha)+ 2m_{c}^{2}\alpha}{(1-\alpha)}
\delta\Big[s-\frac{m_{b}^{2} (1-\alpha)+m_{c}^{2}\alpha}{\alpha(1-\alpha)}\Big]\\
&-2\alpha H\Big[s-\frac{m_{b}^{2} (1-\alpha)+m_{c}^{2}\alpha}{\alpha(1-\alpha)}\Big]\Big\}\, .
\end{aligned}
\end{equation}\\

8. Spectral densities for $J_{4\mu}$:\\
$\rho_{J_{4\mu}}^{0}(s)=\frac{1}{2}\rho_{J_{3\mu}}^{0}(s)\, ,
~\rho_{J_{4\mu}}^{3}(s)=\frac{1}{2}\rho_{J_{3\mu}}^{3}(s)\, ,
~\rho_{J_{4\mu}}^{4a}(s)=\frac{1}{2}\rho_{J_{3\mu}}^{4a}(s)\, ,
~\rho_{J_{4\mu}}^{4b}(s)=-\rho_{J_{3\mu}}^{4b}(s)\, ,
~\rho_{J_{4\mu}}^{5a}(s)=\frac{1}{2}\rho_{J_{3\mu}}^{5a}(s)\, ,
~\rho_{J_{4\mu}}^{5b}(s)=\frac{1}{2}\rho_{J_{3\mu}}^{5b}(s)\, ,
~\rho_{J_{4\mu}}^{6}(s)=\frac{1}{2}\rho_{J_{3\mu}}^{6}(s)\, ,
~\rho_{J_{4\mu}}^{8}(s)=\frac{1}{2}\rho_{J_{3\mu}}^{8}(s)$\, .\\

\subsection{ The spectral densities for $sc\bar{q}\bar{b}$ tetraquarks}
1. Spectral densities for $\eta_{1}$:\\
\begin{equation}\nonumber
\begin{aligned}
\rho_{\eta1}^{0}(s)=&
\int_{\alpha_{min}}^{\alpha_{max}} d \alpha \int_{\beta_{min}}^{\beta_{max}} d \beta
 \frac{(1-\alpha-\beta)^{2}(m_{b}^{2} \beta+m_{c}^{2} \alpha-\alpha \beta s)^{2}} {256 \pi^{6}}
 \Big[\frac{1}{ \alpha^{3} \beta^{3}}(m_{b}^{2} \beta+m_{c}^{2} \alpha-\alpha \beta s)\\
 &\times(m_{b}^{2} \beta+m_{c}^{2} \alpha-3 \alpha \beta s)-\Big(\frac{2m_{b}m_{q}}{\alpha^{3} \beta^{2}}+\frac{2m_{c}m_{s}}{ \alpha^{2} \beta^{3}}\Big)(2m_{b}^{2} \beta+2m_{c}^{2} \alpha-5 \alpha \beta s)\Big]\, ,
 \end{aligned}
\end{equation}
\begin{equation}\nonumber
 \begin{aligned}
\rho_{\eta_{1}}^{3}(s)=&
\Big[m_{q}\langle\bar{q}q\rangle + m_{s}\langle\bar{s}s\rangle \Big]
\int_{\alpha_{min}}^{\alpha_{max}} d \alpha \int_{\beta_{min}}^{\beta_{max}} d \beta
\frac{(m_{b}^{2} \beta+m_{c}^{2} \alpha-\alpha \beta s)
(m_{b}^{2} \beta+m_{c}^{2} \alpha-2 \alpha \beta s)}{16 \pi^{4}\alpha \beta}\\
 &+\Big[m_{s}\langle\bar{q}q\rangle + m_{q}\langle\bar{s}s\rangle \Big]
\int_{\alpha_{min}}^{\alpha_{max}} d \alpha \int_{\beta_{min}}^{\beta_{max}} d \beta
\frac{(m_{b}^{2} \beta+m_{c}^{2} \alpha-\alpha \beta s)m_{b}m_{c}}{8 \pi^{4}\alpha \beta}\\
 &-m_{b}\langle\bar{q}q\rangle
\int_{\alpha_{min}}^{\alpha_{max}} d \alpha \int_{\beta_{min}}^{\beta_{max}} d \beta
\frac{(1-\alpha-\beta)(m_{b}^{2} \beta+m_{c}^{2} \alpha-\alpha \beta s)(m_{b}^{2} \beta+m_{c}^{2} \alpha-2 \alpha \beta s)}{8 \pi^{4}\alpha^{2} \beta}\\
&-m_{c}\langle\bar{s}s\rangle
\int_{\alpha_{min}}^{\alpha_{max}} d \alpha \int_{\beta_{min}}^{\beta_{max}} d \beta
\frac{(1-\alpha-\beta)(m_{b}^{2} \beta+m_{c}^{2} \alpha-\alpha \beta s)(m_{b}^{2} \beta+m_{c}^{2} \alpha-2 \alpha \beta s)}{8 \pi^{4} \alpha \beta^{2}}\, ,
 \end{aligned}
\end{equation}
\begin{equation}\nonumber
\begin{aligned}
\rho_{\eta_{1}}^{4a}(s)=
&\langle g_{s}^{2} G G\rangle \int_{\alpha {min }}^{\alpha_{max }} d \alpha \int_{\beta_{min }}^{\beta_{\max }} d \beta \frac{(1-\alpha-\beta)^{2}(2 m_{b}^{2} \beta+2 m_{c}^{2} \alpha-3 \alpha \beta s)}{1536 \pi^{6}}\Big(\frac{m_{b}^{2}}{\alpha^{3}}+\frac{m_{c}^{2}}{\beta^{3}}\Big)\, ,
\end{aligned}
\end{equation}
\begin{equation}\nonumber
\begin{aligned}
\rho_{\eta_{1}}^{4b}(s)=
&-\langle g_{s}^{2} G G\rangle
\int_{\alpha_{min }}^{\alpha_{max }} d \alpha \int_{\beta_{min }}^{\beta_{max }} d \beta \frac{(1-\alpha-\beta)(m_{b}^{2} \beta+m_{c}^{2} \alpha- \alpha \beta s)(m_{b}^{2} \beta+m_{c}^{2} \alpha-2\alpha \beta s)}{1024 \pi^{6} \alpha \beta}\Big(\frac{1}{\alpha}+\frac{1}{\beta}\Big)\, ,
\end{aligned}
\end{equation}
\begin{equation}\nonumber
\begin{aligned}
\rho_{\eta_{1}}^{5a}(s)=&
 m_{b}\langle\bar{q} g_{s}\sigma\cdot G q\rangle\int_{\alpha_{min }}^{\alpha_{max }} d \alpha \int_{\beta_{min }}^{\beta_{max }} d \beta \frac{2 m_{b}^{2} \beta+2 m_{c}^{2} \alpha-3 \alpha \beta s}{32 \pi^{4}\alpha}\\
&+ m_{c}\langle\bar{s} g_{s}\sigma\cdot G s\rangle\int_{\alpha_{min }}^{\alpha_{max }} d \alpha \int_{\beta_{min }}^{\beta_{max }} d \beta
\frac{2 m_{b}^{2} \beta+2 m_{c}^{2} \alpha-3 \alpha \beta s}{32 \pi^{4}\beta}\, ,
\end{aligned}
\end{equation}
\begin{equation}\nonumber
\begin{aligned}
\rho_{\eta_{1}}^{5b}(s)=&
 \langle\bar{q} g_{s}\sigma\cdot G q\rangle\int_{\alpha_{min }}^{\alpha_{max }} d \alpha \int_{\beta_{min }}^{\beta_{max }} d \beta \Big[\frac{m_{b}(1-\alpha-\beta)(2 m_{b}^{2} \beta+2 m_{c}^{2} \alpha-3 \alpha \beta s)}{64 \pi^{4}\alpha^{2}}-\frac{m_{b}m_{c}m_{s}}{64 \pi^{4}\alpha}\Big]\\
&+ \langle\bar{s} g_{s}\sigma\cdot G s\rangle\int_{\alpha_{min }}^{\alpha_{max }} d \alpha \int_{\beta_{min }}^{\beta_{max }} d \beta
\Big[\frac{m_{c}(1-\alpha-\beta)(2 m_{b}^{2} \beta+2 m_{c}^{2} \alpha-3 \alpha \beta s)}{64 \pi^{4}\beta^{2}}-\frac{m_{b}m_{c}m_{q}}{64 \pi^{4}\beta}\Big]\, ,
\end{aligned}
\end{equation}
\begin{equation}\nonumber
\begin{aligned}
\rho_{\eta_{1}}^{5c}(s)=&
\Big[ m_{q}\langle\bar{q} g_{s}\sigma\cdot G q\rangle+m_{s}\langle\bar{s} g_{s}\sigma\cdot G s\rangle\Big]\frac{s-m_{b}^{2}-m_{c}^{2}}{192 \pi^{4}}\sqrt{\left(1+\frac{m_{b}^{2}-m_{c}^{2}}{s}\right)^{2}-\frac{4 m_{b}^{2}}{s}} \\
&-\Big[ m_{q}\langle\bar{s} g_{s}\sigma\cdot G s\rangle+m_{s}\langle\bar{q} g_{s}\sigma\cdot G q\rangle\Big]\frac{m_{b}m_{c}}{32 \pi^{4}}\sqrt{\left(1+\frac{m_{b}^{2}-m_{c}^{2}}{s}\right)^{2}-\frac{4 m_{b}^{2}}{s}}\, ,
\end{aligned}
\end{equation}
\begin{equation}\nonumber
\begin{aligned}
\rho_{\eta_{1}}^{6}(s)=&
\frac{\langle\bar{q} q\rangle \langle\bar{s} s\rangle}{12 \pi^{2}}\Big[ 2m_{b}m_{c}+m_{b}m_{s}+m_{c}m_{q}-\frac{m_{b}^{2}m_{c}m_{q}(s-m_{b}^{2}+m_{c}^{2})+m_{b}m_{c}^{2}m_{s}(s+m_{b}^{2}-m_{c}^{2})}{(s-m_{b}^{2}+m_{c}^{2})^{2}-4m_{c}^{2}s}\Big]\\
&\times \sqrt{\left(1+\frac{m_{b}^{2}-m_{c}^{2}}{s}\right)^{2}-\frac{4 m_{b}^{2}}{s}}\, ,
\end{aligned}
\end{equation}
\begin{equation}\nonumber
\begin{aligned}
\rho_{\eta_{1}}^{8a}(s)=
&\Big[  \langle\bar{q} q\rangle \langle\bar{s}g_{s}\sigma\cdot G s\rangle+\langle\bar{s} s\rangle \langle\bar{q}g_{s}\sigma\cdot G q\rangle \Big]
 \int_{0}^{1} d \alpha \frac{1}{24 \pi^{2}}
 \frac{ m_{b} m_{c}^{3}}{(1-\alpha)^{2}} \delta^{\prime}\Big[s-\frac{m_{b}^{2} (1-\alpha)+m_{c}^{2}\alpha}{\alpha(1-\alpha)}\Big]\, ,
\end{aligned}
\end{equation}
\begin{equation}\nonumber
\begin{aligned}
\rho_{\eta_{1}}^{8b}(s)=
&  \langle\bar{q} q\rangle \langle\bar{s}g_{s}\sigma\cdot G s\rangle
 \int_{0}^{1} d \alpha
 \frac{ m_{b} m_{c}}{48 \pi^{2}(1-\alpha)} \delta\Big[s-\frac{m_{b}^{2} (1-\alpha)+m_{c}^{2}\alpha}{\alpha(1-\alpha)}\Big]\\
 &+ \langle\bar{s} s\rangle \langle\bar{q}g_{s}\sigma\cdot G q\rangle
 \int_{0}^{1} d \alpha
 \frac{ m_{b} m_{c}}{48 \pi^{2}\alpha} \delta\Big[s-\frac{m_{b}^{2} (1-\alpha)+m_{c}^{2}\alpha}{\alpha(1-\alpha)}\Big]\, .
\end{aligned}
\end{equation}\\

2. Spectral densities for $\eta_{2}$:\\
$\rho_{\eta_{2}}^{0}(s)=\frac{1}{2}\rho_{\eta_{1}}^{0}(s)\, ,
~\rho_{\eta_{2}}^{3}(s)=\frac{1}{2}\rho_{\eta_{1}}^{3}(s)\, ,
~\rho_{\eta_{2}}^{4a}(s)=\frac{1}{2}\rho_{\eta_{1}}^{4a}(s)\, ,
~\rho_{\eta_{2}}^{4b}(s)=-\rho_{\eta_{1}}^{4b}(s)\, ,
~\rho_{\eta_{2}}^{5a}(s)=\frac{1}{2}\rho_{\eta_{1}}^{5a}(s)\, ,
~\rho_{\eta_{2}}^{5b}(s)=-\rho_{\eta_{1}}^{5b}(s)\, ,
~\rho_{\eta_{2}}^{5c}(s)=\frac{1}{2}\rho_{\eta_{1}}^{5c}(s)\, ,
~\rho_{\eta_{2}}^{6}(s)=\frac{1}{2}\rho_{\eta_{1}}^{6}(s)\, ,
~\rho_{\eta_{2}}^{8a}(s)=\frac{1}{2}\rho_{\eta_{1}}^{8a}(s)\, ,
~\rho_{\eta_{2}}^{8b}(s)=-\rho_{\eta_{1}}^{8b}(s)$\, .\\

3. Spectral densities for $\eta_{3}$:\\
\begin{equation}\nonumber
\begin{aligned}
\rho_{\eta_{3}}^{0}(s)=&
\int_{\alpha_{min}}^{\alpha_{max}} d \alpha \int_{\beta_{min}}^{\beta_{max}} d \beta
 \frac{(1-\alpha-\beta)^{2}(m_{b}^{2} \beta+m_{c}^{2} \alpha-\alpha \beta s)^{2}} {64 \pi^{6}}
 \Big[\frac{1}{ \alpha^{3} \beta^{3}}(m_{b}^{2} \beta+m_{c}^{2} \alpha-\alpha \beta s)\\
 &\times(m_{b}^{2} \beta+m_{c}^{2} \alpha-3 \alpha \beta s)-\Big(\frac{m_{b}m_{q}}{\alpha^{3} \beta^{2}}+\frac{m_{c}m_{s}}{ \alpha^{2} \beta^{3}}\Big)(2m_{b}^{2} \beta+2m_{c}^{2} \alpha-5 \alpha \beta s)\Big]\, ,
 \end{aligned}
\end{equation}
\begin{equation}\nonumber
 \begin{aligned}
\rho_{\eta_{3}}^{3}(s)=&
\Big[m_{q}\langle\bar{q}q\rangle + m_{s}\langle\bar{s}s\rangle \Big]
\int_{\alpha_{min}}^{\alpha_{max}} d \alpha \int_{\beta_{min}}^{\beta_{max}} d \beta
\frac{(m_{b}^{2} \beta+m_{c}^{2} \alpha-\alpha \beta s)
(m_{b}^{2} \beta+m_{c}^{2} \alpha-2 \alpha \beta s)}{4 \pi^{4}\alpha \beta}\\
 &+\Big[m_{s}\langle\bar{q}q\rangle + m_{q}\langle\bar{s}s\rangle \Big]
\int_{\alpha_{min}}^{\alpha_{max}} d \alpha \int_{\beta_{min}}^{\beta_{max}} d \beta
\frac{(m_{b}^{2} \beta+m_{c}^{2} \alpha-\alpha \beta s)m_{b}m_{c}}{2 \pi^{4}\alpha \beta}\\
 &-m_{b}\langle\bar{q}q\rangle
\int_{\alpha_{min}}^{\alpha_{max}} d \alpha \int_{\beta_{min}}^{\beta_{max}} d \beta
\frac{(1-\alpha-\beta)(m_{b}^{2} \beta+m_{c}^{2} \alpha-\alpha \beta s)(m_{b}^{2} \beta+m_{c}^{2} \alpha-2 \alpha \beta s)}{4 \pi^{4}\alpha^{2} \beta}\\
&-m_{c}\langle\bar{s}s\rangle
\int_{\alpha_{min}}^{\alpha_{max}} d \alpha \int_{\beta_{min}}^{\beta_{max}} d \beta
\frac{(1-\alpha-\beta)(m_{b}^{2} \beta+m_{c}^{2} \alpha-\alpha \beta s)(m_{b}^{2} \beta+m_{c}^{2} \alpha-2 \alpha \beta s)}{4 \pi^{4} \alpha \beta^{2}}\, ,
 \end{aligned}
\end{equation}
\begin{equation}\nonumber
\begin{aligned}
\rho_{\eta_{3}}^{4a}(s)=
&\langle g_{s}^{2} G G\rangle \int_{\alpha {min }}^{\alpha_{max }} d \alpha \int_{\beta_{min }}^{\beta_{\max }} d \beta \frac{(1-\alpha-\beta)^{2}(2 m_{b}^{2} \beta+2 m_{c}^{2} \alpha-3 \alpha \beta s)}{384 \pi^{6}}\Big(\frac{m_{b}^{2}}{\alpha^{3}}+\frac{m_{c}^{2}}{\beta^{3}}\Big)\, ,
\end{aligned}
\end{equation}
\begin{equation}\nonumber
\begin{aligned}
\rho_{\eta_{3}}^{4b}(s)=
&\langle g_{s}^{2} G G\rangle
\int_{\alpha_{min }}^{\alpha_{max }} d \alpha \int_{\beta_{min }}^{\beta_{max }} d \beta \frac{(1-\alpha-\beta)(m_{b}^{2} \beta+m_{c}^{2} \alpha-\alpha \beta s)(m_{b}^{2} \beta+m_{c}^{2} \alpha-2\alpha \beta s)}{512 \pi^{6} \alpha \beta}\Big(\frac{1}{\alpha}+\frac{1}{\beta}\Big)\, ,
\end{aligned}
\end{equation}
\begin{equation}\nonumber
\begin{aligned}
\rho_{\eta_{3}}^{4c}(s)=
&\langle g_{s}^{2} G G\rangle
\int_{\alpha_{min }}^{\alpha_{max }} d \alpha \int_{\beta_{min }}^{\beta_{max }} d \beta \frac{(m_{b}^{2} \beta+m_{c}^{2} \alpha-\alpha \beta s)(m_{b}^{2} \beta+m_{c}^{2} \alpha-2\alpha \beta s)}{512 \pi^{6} \alpha \beta}\Big[1+\frac{(1-\alpha-\beta)^{2}}{2\alpha\beta}\Big]\, ,
\end{aligned}
\end{equation}
\begin{equation}\nonumber
\begin{aligned}
\rho_{\eta_{3}}^{5a}(s)=&
 m_{b}\langle\bar{q} g_{s}\sigma\cdot G q\rangle\int_{\alpha_{min }}^{\alpha_{max }} d \alpha \int_{\beta_{min }}^{\beta_{max }} d \beta \frac{2 m_{b}^{2} \beta+2 m_{c}^{2} \alpha-3 \alpha \beta s}{16 \pi^{4}\alpha}\\
&+ m_{c}\langle\bar{s} g_{s}\sigma\cdot G s\rangle\int_{\alpha_{min }}^{\alpha_{max }} d \alpha \int_{\beta_{min }}^{\beta_{max }} d \beta
\frac{2 m_{b}^{2} \beta+2 m_{c}^{2} \alpha-3 \alpha \beta s}{16\pi^{4}\beta}\, ,
\end{aligned}
\end{equation}
\begin{equation}\nonumber
\begin{aligned}
\rho_{\eta_{3}}^{5b}(s)=&
\Big[ m_{q}\langle\bar{q} g_{s}\sigma\cdot G q\rangle+m_{s}\langle\bar{s} g_{s}\sigma\cdot G s\rangle\Big]\frac{s-m_{b}^{2}-m_{c}^{2}}{48 \pi^{4}}\sqrt{\left(1+\frac{m_{b}^{2}-m_{c}^{2}}{s}\right)^{2}-\frac{4 m_{b}^{2}}{s}} \\
&-\Big[ m_{q}\langle\bar{s} g_{s}\sigma\cdot G s\rangle+m_{s}\langle\bar{q} g_{s}\sigma\cdot G q\rangle\Big]\frac{m_{b}m_{c}}{8 \pi^{4}}\sqrt{\left(1+\frac{m_{b}^{2}-m_{c}^{2}}{s}\right)^{2}-\frac{4 m_{b}^{2}}{s}}\, ,
\end{aligned}
\end{equation}
\begin{equation}\nonumber
\begin{aligned}
\rho_{\eta_{3}}^{6}(s)=&
\frac{\langle\bar{q} q\rangle \langle\bar{s} s\rangle}{6 \pi^{2}}\Big[ 4m_{b}m_{c}+m_{b}m_{s}+m_{c}m_{q}-\frac{m_{b}^{2}m_{c}m_{q}(s-m_{b}^{2}+m_{c}^{2})+m_{b}m_{c}^{2}m_{s}(s+m_{b}^{2}-m_{c}^{2})}{(s-m_{b}^{2}+m_{c}^{2})^{2}-4m_{c}^{2}s}\Big]\\
&\times \sqrt{\left(1+\frac{m_{b}^{2}-m_{c}^{2}}{s}\right)^{2}-\frac{4 m_{b}^{2}}{s}}\, ,
\end{aligned}
\end{equation}
\begin{equation}\nonumber
\begin{aligned}
\rho_{\eta_{3}}^{8a}(s)=
&\Big[  \langle\bar{q} q\rangle \langle\bar{s}g_{s}\sigma\cdot G s\rangle+\langle\bar{s} s\rangle \langle\bar{q}g_{s}\sigma\cdot G q\rangle \Big]
 \int_{0}^{1} d \alpha
 \frac{ m_{b} m_{c}^{3}}{6 \pi^{2}(1-\alpha)^{2}} \delta^{\prime}\Big[s-\frac{m_{b}^{2} (1-\alpha)+m_{c}^{2}\alpha}{\alpha(1-\alpha)}\Big]\, .
\end{aligned}
\end{equation}\\

4. Spectral densities for $\eta_{4}$:\\
$\rho_{\eta_{4}}^{0}(s)=\frac{1}{2}\rho_{\eta_{3}}^{0}(s)\, ,
~\rho_{\eta_{4}}^{3}(s)=\frac{1}{2}\rho_{\eta_{3}}^{3}(s)\, ,
~\rho_{\eta_{4}}^{4a}(s)=\frac{1}{2}\rho_{\eta_{3}}^{4a}(s)\, ,
~\rho_{\eta_{4}}^{4b}(s)=-\rho_{\eta_{3}}^{4b}(s)\, ,
~\rho_{\eta_{4}}^{4c}(s)=-\rho_{\eta_{3}}^{4c}(s)\, ,
~\rho_{\eta_{4}}^{5a}(s)=\frac{1}{2}\rho_{\eta_{3}}^{5a}(s)\, ,
~\rho_{\eta_{4}}^{5b}(s)=\frac{1}{2}\rho_{\eta_{3}}^{5b}(s)\, ,
~\rho_{\eta_{4}}^{6}(s)=\frac{1}{2}\rho_{\eta_{3}}^{6}(s)\, ,
~\rho_{\eta_{4}}^{8a}(s)=\frac{1}{2}\rho_{\eta_{3}}^{8a}(s)$\, .\\

5. Spectral densities for $\eta_{1\mu}$:\\
\begin{equation}\nonumber
\begin{aligned}
\rho_{\eta_{1\mu}}^{0}(s)=&
\int_{\alpha_{min}}^{\alpha_{max}} d \alpha \int_{\beta_{min}}^{\beta_{max}} d \beta
 \frac{(1-\alpha-\beta)^{2}(m_{b}^{2} \beta+m_{c}^{2} \alpha-\alpha \beta s)^{2}}{128\pi^{6} \alpha^{2}\beta^{2}}
  \Big\{
 \frac{1}{12\alpha\beta}\Big[6(m_{b}^{2} \beta+m_{c}^{2} \alpha-\alpha \beta s)\\
 &\times(m_{b}^{2} \beta+m_{c}^{2} \alpha-3 \alpha \beta s)-(1-\alpha-\beta)(m_{b}^{2} \beta+m_{c}^{2} \alpha-\alpha \beta s)(3m_{b}^{2} \beta+3m_{c}^{2} \alpha-7 \alpha \beta s)\Big]\\
 &-
 \Big[\frac{(2m_{b}^{2} \beta+2m_{c}^{2} \alpha-5\alpha \beta s)m_{b}m_{q}}{\alpha}+\frac{(m_{b}^{2} \beta+m_{c}^{2} \alpha-4 \alpha \beta s)m_{c}m_{s}}{\beta}\Big]\Big\}\, ,
 \end{aligned}
\end{equation}
\begin{equation}\nonumber
 \begin{aligned}
\rho_{\eta_{1\mu}}^{3}(s)=&
\Big[m_{q}\langle\bar{q}q\rangle + m_{s}\langle\bar{s}s\rangle \Big]
\int_{\alpha_{min}}^{\alpha_{max}} d \alpha \int_{\beta_{min}}^{\beta_{max}} d \beta
\frac{(m_{b}^{2} \beta+m_{c}^{2} \alpha-\alpha \beta s)}{32 \pi^{4}\alpha \beta}
\big[2(m_{b}^{2} \beta+m_{c}^{2} \alpha-2 \alpha \beta s)\\
&-(1-\alpha-\beta)(3m_{b}^{2} \beta+3m_{c}^{2} \alpha-5 \alpha \beta s)\big]\\
 &+\Big[m_{s}\langle\bar{q}q\rangle + m_{q}\langle\bar{s}s\rangle \Big]
\int_{\alpha_{min}}^{\alpha_{max}} d \alpha \int_{\beta_{min}}^{\beta_{max}} d \beta
\frac{(m_{b}^{2} \beta+m_{c}^{2} \alpha-\alpha \beta s)m_{b}m_{c}}{8 \pi^{4}\alpha \beta}\\
 &-m_{b}\langle\bar{q}q\rangle
\int_{\alpha_{min}}^{\alpha_{max}} d \alpha \int_{\beta_{min}}^{\beta_{max}} d \beta
\frac{(1-\alpha-\beta)(m_{b}^{2} \beta+m_{c}^{2} \alpha-\alpha \beta s)(m_{b}^{2} \beta+m_{c}^{2} \alpha-2 \alpha \beta s)}{8 \pi^{4}\alpha^{2} \beta}\\
&-m_{c}\langle\bar{s}s\rangle
\int_{\alpha_{min}}^{\alpha_{max}} d \alpha \int_{\beta_{min}}^{\beta_{max}} d \beta
\frac{(1-\alpha-\beta)(m_{b}^{2} \beta+m_{c}^{2} \alpha-\alpha \beta s)(m_{b}^{2} \beta+m_{c}^{2} \alpha-3 \alpha \beta s)}{16 \pi^{4} \alpha \beta^{2}}\, ,
 \end{aligned}
\end{equation}
\begin{equation}\nonumber
\begin{aligned}
\rho_{\eta_{1\mu}}^{4a}(s)=
&\langle g_{s}^{2} G G\rangle \int_{\alpha {min }}^{\alpha_{max }} d \alpha \int_{\beta_{min }}^{\beta_{\max }} d \beta
\frac{(1-\alpha-\beta)^{2}}{4608\pi^{6}}[3(2 m_{b}^{2} \beta+2 m_{c}^{2} \alpha-3 \alpha \beta s)\\
&-(1-\alpha-\beta)(3 m_{b}^{2} \beta+3 m_{c}^{2} \alpha-4 \alpha \beta s)]\Big(\frac{m_{b}^{2}}{\alpha^{3}}+\frac{m_{c}^{2}}{\beta^{3}}\Big)\, ,
\end{aligned}
\end{equation}
\begin{equation}\nonumber
\begin{aligned}
\rho_{\eta_{1\mu}}^{4b}(s)=
&-\langle g_{s}^{2} G G\rangle
\int_{\alpha_{min }}^{\alpha_{max }} d \alpha \int_{\beta_{min }}^{\beta_{max }} d \beta
\frac{(1-\alpha-\beta)(m_{b}^{2}\beta+m_{c}^{2}\alpha-\alpha \beta s)}{4096 \pi^{6} \alpha \beta}
\Big[\frac{4(3\alpha-\beta)(m_{b}^{2} \beta+m_{c}^{2} \alpha-2 \alpha \beta s)}{3\alpha\beta}\\
&-\frac{(3\alpha+\beta)(1-\alpha-\beta)(3m_{b}^{2} \beta+3m_{c}^{2} \alpha-5 \alpha \beta s)}{3\alpha\beta}\Big]\, ,
\end{aligned}
\end{equation}
\begin{equation}\nonumber
\begin{aligned}
\rho_{\eta_{1\mu}}^{5a}(s)=&
 \langle\bar{q} g_{s}\sigma\cdot G q\rangle\int_{\alpha_{min }}^{\alpha_{max }} d \alpha \int_{\beta_{min }}^{\beta_{max }} d \beta
 \Big[\frac{m_{b}(2 m_{b}^{2} \beta+2 m_{c}^{2} \alpha-3 \alpha \beta s)}{32 \pi^{4}\alpha}+\frac{m_{q}(3 m_{b}^{2} \beta+3 m_{c}^{2} \alpha-4 \alpha \beta s)}{96 \pi^{4}}\Big]\\
&+\langle\bar{s} g_{s}\sigma\cdot G s\rangle\int_{\alpha_{min }}^{\alpha_{max }} d \alpha \int_{\beta_{min }}^{\beta_{max }} d \beta
\Big[\frac{  m_{c}(m_{b}^{2} \beta+ m_{c}^{2} \alpha-2 \alpha \beta s)}{32 \pi^{4}\beta}+\frac{m_{s}(3 m_{b}^{2} \beta+3 m_{c}^{2} \alpha-4 \alpha \beta s)}{96 \pi^{4}}\Big]\, ,
\end{aligned}
\end{equation}
\begin{equation}\nonumber
\begin{aligned}
\rho_{\eta_{1\mu}}^{5b}(s)=&
 \langle\bar{s} g_{s}\sigma\cdot G s\rangle\int_{\alpha_{min }}^{\alpha_{max }} d \alpha \int_{\beta_{min }}^{\beta_{max }} d \beta
\Big[\frac{m_{c}(1-\alpha-\beta)( m_{b}^{2} \beta+ m_{c}^{2} \alpha-2 \alpha \beta s)}{64 \pi^{4}\beta^{2}}-\frac{m_{b}m_{c}m_{q}}{64 \pi^{4}\beta}\Big]\, ,
\end{aligned}
\end{equation}
\begin{equation}\nonumber
\begin{aligned}
\rho_{\eta_{1\mu}}^{5c}(s)=&
\Big[ m_{q}\langle\bar{q} g_{s}\sigma\cdot G q\rangle+m_{s}\langle\bar{s} g_{s}\sigma\cdot G s\rangle\Big]\frac{s-m_{b}^{2}-m_{c}^{2}}{192 \pi^{4}}\sqrt{\left(1+\frac{m_{b}^{2}-m_{c}^{2}}{s}\right)^{2}-\frac{4 m_{b}^{2}}{s}} \\
&-\Big[ m_{q}\langle\bar{s} g_{s}\sigma\cdot G s\rangle+m_{s}\langle\bar{q} g_{s}\sigma\cdot G q\rangle\Big]\frac{m_{b}m_{c}}{32 \pi^{4}}\sqrt{\left(1+\frac{m_{b}^{2}-m_{c}^{2}}{s}\right)^{2}-\frac{4 m_{b}^{2}}{s}}\, ,
\end{aligned}
\end{equation}
\begin{equation}\nonumber
\begin{aligned}
\rho_{\eta_{1\mu}}^{6}(s)=&
\frac{\langle\bar{q} q\rangle \langle\bar{s} s\rangle}{12 \pi^{2}}\Big[ 2m_{b}m_{c}+m_{b}m_{s}+m_{c}m_{q}-\frac{m_{b}^{2}m_{c}m_{q}(s-m_{b}^{2}+m_{c}^{2})+m_{b}m_{c}^{2}m_{s}(s+m_{b}^{2}-m_{c}^{2})}{(s-m_{b}^{2}+m_{c}^{2})^{2}-4m_{c}^{2}s}\\
&-\frac{m_{c}m_{q}(s+m_{b}^{2}-m_{c}^{2})}{2s}\Big]\sqrt{\left(1+\frac{m_{b}^{2}-m_{c}^{2}}{s}\right)^{2}-\frac{4 m_{b}^{2}}{s}}\, ,
\end{aligned}
\end{equation}
\begin{equation}\nonumber
\begin{aligned}
\rho_{\eta_{1\mu}}^{8a}(s)=
&\Big[  \langle\bar{q} q\rangle \langle\bar{s}g_{s}\sigma\cdot G s\rangle+\langle\bar{s} s\rangle \langle\bar{q}g_{s}\sigma\cdot G q\rangle \Big]
 \int_{0}^{1} d \alpha \frac{1}{24 \pi^{2}}
 \frac{ m_{b} m_{c}^{3}}{(1-\alpha)^{2}} \delta^{\prime}\Big[s-\frac{m_{b}^{2} (1-\alpha)+m_{c}^{2}\alpha}{\alpha(1-\alpha)}\Big]\, ,
\end{aligned}
\end{equation}
\begin{equation}\nonumber
\begin{aligned}
\rho_{\eta_{1\mu}}^{8b}(s)=
&  \langle\bar{q} q\rangle \langle\bar{s}g_{s}\sigma\cdot G s\rangle
 \int_{0}^{1} d \alpha
 \frac{ m_{b} m_{c}}{48 \pi^{2}(1-\alpha)} \delta\Big[s-\frac{m_{b}^{2} (1-\alpha)+m_{c}^{2}\alpha}{\alpha(1-\alpha)}\Big]\, .
\end{aligned}
\end{equation}\\

6. Spectral densities for $\eta_{2\mu}$:\\
$\rho_{\eta_{2\mu}}^{0}(s)=\frac{1}{2}\rho_{\eta_{1\mu}}^{0}(s)\, ,
~\rho_{\eta_{2\mu}}^{3}(s)=\frac{1}{2}\rho_{\eta_{1\mu}}^{3}(s)\, ,
~\rho_{\eta_{2\mu}}^{4a}(s)=\frac{1}{2}\rho_{\eta_{1\mu}}^{4a}(s)\, ,
~\rho_{\eta_{2\mu}}^{4b}(s)=-\rho_{\eta_{1\mu}}^{4b}(s)\, ,
~\rho_{\eta_{2\mu}}^{5a}(s)=\frac{1}{2}\rho_{\eta_{1\mu}}^{5a}(s)\, ,
~\rho_{\eta_{2\mu}}^{5b}(s)=-\rho_{\eta_{1\mu}}^{5b}(s)\, ,
~\rho_{\eta_{2\mu}}^{5c}(s)=\frac{1}{2}\rho_{\eta_{1\mu}}^{5c}(s)\, ,
~\rho_{\eta_{2\mu}}^{6}(s)=\frac{1}{2}\rho_{\eta_{1\mu}}^{6}(s)\, ,
~\rho_{\eta_{2\mu}}^{8a}(s)=\frac{1}{2}\rho_{\eta_{1\mu}}^{8a}(s)\, ,
~\rho_{\eta_{2\mu}}^{8b}(s)=-\rho_{\eta_{1\mu}}^{8b}(s)$\, .\\

7. Spectral densities for $\eta_{3\mu}$;\\
\begin{equation}\nonumber
\begin{aligned}
\rho_{\eta_{3\mu}}^{0}(s)=&
\int_{\alpha_{min}}^{\alpha_{max}} d \alpha \int_{\beta_{min}}^{\beta_{max}} d \beta
 \frac{(1-\alpha-\beta)^{2}(m_{b}^{2} \beta+m_{c}^{2} \alpha-\alpha \beta s)^{2}}{128\pi^{6} \alpha^{2}\beta^{2}}
  \Big\{
 \frac{1}{12\alpha\beta}\Big[6(m_{b}^{2} \beta+m_{c}^{2} \alpha-\alpha \beta s)\\
 &\times(m_{b}^{2} \beta+m_{c}^{2} \alpha-3 \alpha \beta s)-(1-\alpha-\beta)(m_{b}^{2} \beta+m_{c}^{2} \alpha-\alpha \beta s)(3m_{b}^{2} \beta+3m_{c}^{2} \alpha-7 \alpha \beta s)\Big]\\
 &-
 \Big[\frac{(m_{b}^{2} \beta+m_{c}^{2} \alpha-4\alpha \beta s)m_{b}m_{q}}{\alpha}+\frac{(2m_{b}^{2} \beta+2m_{c}^{2} \alpha-5\alpha \beta s)m_{c}m_{s}}{\beta}\Big]\Big\}\, ,
 \end{aligned}
\end{equation}
\begin{equation}\nonumber
 \begin{aligned}
\rho_{\eta_{3\mu}}^{3}(s)=&
\Big[m_{q}\langle\bar{q}q\rangle + m_{s}\langle\bar{s}s\rangle \Big]
\int_{\alpha_{min}}^{\alpha_{max}} d \alpha \int_{\beta_{min}}^{\beta_{max}} d \beta
\frac{(m_{b}^{2} \beta+m_{c}^{2} \alpha-\alpha \beta s)}{32 \pi^{4}\alpha \beta}
\big[2(m_{b}^{2} \beta+m_{c}^{2} \alpha-2 \alpha \beta s)\\
&-(1-\alpha-\beta)(3m_{b}^{2} \beta+3m_{c}^{2} \alpha-5 \alpha \beta s)\big]\\
 &+\Big[m_{s}\langle\bar{q}q\rangle + m_{q}\langle\bar{s}s\rangle \Big]
\int_{\alpha_{min}}^{\alpha_{max}} d \alpha \int_{\beta_{min}}^{\beta_{max}} d \beta
\frac{(m_{b}^{2} \beta+m_{c}^{2} \alpha-\alpha \beta s)m_{b}m_{c}}{8 \pi^{4}\alpha \beta}\\
 &-m_{b}\langle\bar{q}q\rangle
\int_{\alpha_{min}}^{\alpha_{max}} d \alpha \int_{\beta_{min}}^{\beta_{max}} d \beta
\frac{(1-\alpha-\beta)(m_{b}^{2} \beta+m_{c}^{2} \alpha-\alpha \beta s)(m_{b}^{2} \beta+m_{c}^{2} \alpha-3 \alpha \beta s)}{16 \pi^{4}\alpha^{2} \beta}\\
&-m_{c}\langle\bar{s}s\rangle
\int_{\alpha_{min}}^{\alpha_{max}} d \alpha \int_{\beta_{min}}^{\beta_{max}} d \beta
\frac{(1-\alpha-\beta)(m_{b}^{2} \beta+m_{c}^{2} \alpha-\alpha \beta s)(m_{b}^{2} \beta+m_{c}^{2} \alpha-2 \alpha \beta s)}{8 \pi^{4} \alpha \beta^{2}}\, ,
 \end{aligned}
\end{equation}
\begin{equation}\nonumber
\begin{aligned}
\rho_{\eta_{3\mu}}^{4a}(s)=
&\langle g_{s}^{2} G G\rangle \int_{\alpha {min }}^{\alpha_{max }} d \alpha \int_{\beta_{min }}^{\beta_{\max }} d \beta
\frac{(1-\alpha-\beta)^{2}}{4608\pi^{6}}[3(2 m_{b}^{2} \beta+2 m_{c}^{2} \alpha-3 \alpha \beta s)\\
&-(1-\alpha-\beta)(3 m_{b}^{2} \beta+3 m_{c}^{2} \alpha-4 \alpha \beta s)]\Big(\frac{m_{b}^{2}}{\alpha^{3}}+\frac{m_{c}^{2}}{\beta^{3}}\Big)\, ,
\end{aligned}
\end{equation}
\begin{equation}\nonumber
\begin{aligned}
\rho_{\eta_{3\mu}}^{4b}(s)=
&-\langle g_{s}^{2} G G\rangle
\int_{\alpha_{min }}^{\alpha_{max }} d \alpha \int_{\beta_{min }}^{\beta_{max }} d \beta
\frac{(1-\alpha-\beta)(m_{b}^{2}\beta+m_{c}^{2}\alpha-\alpha \beta s)}{4096 \pi^{6} \alpha \beta}
\Big[\frac{4(3\beta-\alpha)(m_{b}^{2} \beta+m_{c}^{2} \alpha-2 \alpha \beta s)}{3\alpha\beta}\\
&-\frac{(\alpha+3\beta)(1-\alpha-\beta)(3m_{b}^{2} \beta+3m_{c}^{2} \alpha-5 \alpha \beta s)}{3\alpha\beta}\Big]\, ,
\end{aligned}
\end{equation}
\begin{equation}\nonumber
\begin{aligned}
\rho_{\eta_{3\mu}}^{5a}(s)=&
 \langle\bar{q} g_{s}\sigma\cdot G q\rangle\int_{\alpha_{min }}^{\alpha_{max }} d \alpha \int_{\beta_{min }}^{\beta_{max }} d \beta
 \Big[\frac{m_{b}( m_{b}^{2} \beta+ m_{c}^{2} \alpha-2 \alpha \beta s)}{32 \pi^{4}\alpha}+\frac{m_{q}(3 m_{b}^{2} \beta+3 m_{c}^{2} \alpha-4 \alpha \beta s)}{96 \pi^{4}}\Big]\\
&+\langle\bar{s} g_{s}\sigma\cdot G s\rangle\int_{\alpha_{min }}^{\alpha_{max }} d \alpha \int_{\beta_{min }}^{\beta_{max }} d \beta
\Big[\frac{  m_{c}(2m_{b}^{2} \beta+ 2m_{c}^{2} \alpha-3 \alpha \beta s)}{32 \pi^{4}\beta}+\frac{m_{s}(3 m_{b}^{2} \beta+3 m_{c}^{2} \alpha-4 \alpha \beta s)}{96 \pi^{4}}\Big]\, ,
\end{aligned}
\end{equation}
\begin{equation}\nonumber
\begin{aligned}
\rho_{\eta_{3\mu}}^{5b}(s)=&
 \langle\bar{q} g_{s}\sigma\cdot G q\rangle\int_{\alpha_{min }}^{\alpha_{max }} d \alpha \int_{\beta_{min }}^{\beta_{max }} d \beta
\Big[\frac{m_{b}(1-\alpha-\beta)( m_{b}^{2} \beta+ m_{c}^{2} \alpha-2 \alpha \beta s)}{64 \pi^{4}\alpha^{2}}-\frac{m_{b}m_{c}m_{s}}{64 \pi^{4}\alpha}\Big]\, ,
\end{aligned}
\end{equation}
\begin{equation}\nonumber
\begin{aligned}
\rho_{\eta_{3\mu}}^{5c}(s)=&
\Big[ m_{q}\langle\bar{q} g_{s}\sigma\cdot G q\rangle+m_{s}\langle\bar{s} g_{s}\sigma\cdot G s\rangle\Big]\frac{s-m_{b}^{2}-m_{c}^{2}}{192 \pi^{4}}\sqrt{\left(1+\frac{m_{b}^{2}-m_{c}^{2}}{s}\right)^{2}-\frac{4 m_{b}^{2}}{s}} \\
&-\Big[ m_{q}\langle\bar{s} g_{s}\sigma\cdot G s\rangle+m_{s}\langle\bar{q} g_{s}\sigma\cdot G q\rangle\Big]\frac{m_{b}m_{c}}{32 \pi^{4}}\sqrt{\left(1+\frac{m_{b}^{2}-m_{c}^{2}}{s}\right)^{2}-\frac{4 m_{b}^{2}}{s}}\, ,
\end{aligned}
\end{equation}
\begin{equation}\nonumber
\begin{aligned}
\rho_{\eta_{3\mu}}^{6}(s)=&
\frac{\langle\bar{q} q\rangle \langle\bar{s} s\rangle}{12 \pi^{2}}\Big[ 2m_{b}m_{c}+m_{b}m_{s}+m_{c}m_{q}-\frac{m_{b}^{2}m_{c}m_{q}(s-m_{b}^{2}+m_{c}^{2})+m_{b}m_{c}^{2}m_{s}(s+m_{b}^{2}-m_{c}^{2})}{(s-m_{b}^{2}+m_{c}^{2})^{2}-4m_{c}^{2}s}\\
&-\frac{m_{b}m_{s}(s-m_{b}^{2}+m_{c}^{2})}{2s}\Big]\sqrt{\left(1+\frac{m_{b}^{2}-m_{c}^{2}}{s}\right)^{2}-\frac{4 m_{b}^{2}}{s}}\, ,
\end{aligned}
\end{equation}
\begin{equation}\nonumber
\begin{aligned}
\rho_{\eta_{3\mu}}^{8a}(s)=
&\Big[  \langle\bar{q} q\rangle \langle\bar{s}g_{s}\sigma\cdot G s\rangle+\langle\bar{s} s\rangle \langle\bar{q}g_{s}\sigma\cdot G q\rangle \Big]
 \int_{0}^{1} d \alpha \frac{1}{24 \pi^{2}}
 \frac{ m_{b} m_{c}^{3}}{(1-\alpha)^{2}} \delta^{\prime}\Big[s-\frac{m_{b}^{2} (1-\alpha)+m_{c}^{2}\alpha}{\alpha(1-\alpha)}\Big]\, ,
\end{aligned}
\end{equation}
\begin{equation}\nonumber
\begin{aligned}
\rho_{\eta_{3\mu}}^{8b}(s)=
&  \langle\bar{s} s\rangle \langle\bar{q}g_{s}\sigma\cdot G q\rangle
 \int_{0}^{1} d \alpha
 \frac{ m_{b} m_{c}}{48 \pi^{2}\alpha} \delta\Big[s-\frac{m_{b}^{2} (1-\alpha)+m_{c}^{2}\alpha}{\alpha(1-\alpha)}\Big]\, .
\end{aligned}
\end{equation}\\

8. Spectral densities for $\eta_{4\mu}$:\\
$\rho_{\eta_{4\mu}}^{0}(s)=\frac{1}{2}\rho_{\eta_{3\mu}}^{0}(s)\, ,
~\rho_{\eta_{4\mu}}^{3}(s)=\frac{1}{2}\rho_{\eta_{3\mu}}^{3}(s)\, ,
~\rho_{\eta_{4\mu}}^{4a}(s)=\frac{1}{2}\rho_{\eta_{3\mu}}^{4a}(s)\, ,
~\rho_{\eta_{4\mu}}^{4b}(s)=-\rho_{\eta_{3\mu}}^{4b}(s)\, ,
~\rho_{\eta_{4\mu}}^{5a}(s)=\frac{1}{2}\rho_{\eta_{3\mu}}^{5a}(s)\, ,
~\rho_{\eta_{4\mu}}^{5b}(s)=-\rho_{\eta_{3\mu}}^{5b}(s)\, ,
~\rho_{\eta_{4\mu}}^{5c}(s)=\frac{1}{2}\rho_{\eta_{3\mu}}^{5c}(s)\, ,
~\rho_{\eta_{4\mu}}^{6}(s)=\frac{1}{2}\rho_{\eta_{3\mu}}^{6}(s)\, ,
~\rho_{\eta_{4\mu}}^{8a}(s)=\frac{1}{2}\rho_{\eta_{3\mu}}^{8a}(s)\, ,
~\rho_{\eta_{4\mu}}^{8b}(s)=-\rho_{\eta_{3\mu}}^{8b}(s)$\, .

%\bibliographystyle{paper}
%\bibliography{paperRef}
\end{document}